\documentclass[nofootinbib,twocolumn]{revtex4-1}
\usepackage{graphicx}
\usepackage{dcolumn}
\usepackage{bm}
\usepackage{helvet} \usepackage{amsmath} \usepackage{graphicx} \usepackage{textcomp} \usepackage[labelfont={bf}]{caption} \usepackage[font={it}]{caption} \usepackage{url}  \usepackage[round-precision=3,round-mode=figures,scientific-notation=true]{siunitx} \usepackage{parskip} \usepackage{mathtools} \usepackage{enumitem} \usepackage{epstopdf} \usepackage{subfig} \usepackage{multirow} \usepackage{morefloats} 
\usepackage{array} \usepackage{bigstrut}
\usepackage{MnSymbol}%
\usepackage{wasysym}%
\usepackage{hyperref}

\makeatletter
\newcommand{\setfoot}[2]{
	\stepcounter{footnote}%
	\expandafter\xdef\csname fn@#1\endcsname{\thefootnAote}
	\protected@xdef\@thanks{\@thanks
		\protect\footnotetext[\the\c@footnote]{#2}}%
	\ignorespaces}
\newcommand{\getfoot}[1]
{\@ifundefined{fn@#1}{?}{\textsuperscript{\csname fn@#1\endcsname}}}
\makeatother

\newcolumntype{C}[1]{>{\centering\let\newline\\\arraybackslash\hspace{0pt}}m{#1}}
\usepackage{indentfirst}
\begin{document}
	\title[Wing optimisation for tractor propeller driven systems in MAVs]{Wing Optimisation for a tractor propeller driven Micro Aerial Vehicle}
	\author{Arjun Sharma}
	\affiliation{
		Sibley School of Mechanical and Aerospace Engineering, Cornell University, Ithaca, NY, 14853, USA 
	}
	\email{as3833@cornell.edu}
	\author{Roddam Narasimha }
\affiliation{
	Jawaharlal Nehru Centre for Advanced Scientific Research, Bengaluru 560064, India 
}
\email{RN passed away on 14 December, 2020. This work stands as a testament to his invaluable contribution and dedication. This manuscript was finished in June, 2018 and AS performed this work primarily with and under the supervision of RN.}
\begin{abstract}
	This paper describes an investigation of the possible benefits from wing optimisation in improving the performance of Micro Air Vehicles (MAVs).  As an example we study the Avion (3.64 kg mass, 1.60 m span), being designed at the CSIR National Aerospace Laboratories (NAL), Bengaluru.  The optimisation is first carried out using the methodology described by Rakshith \emph{et al.} \cite{Rakshith2015} (using an in\textendash house software PROWING), developed for large transport aircraft, with certain modifications to adapt the code to the special features of the MAV.   The chief among such features is the use of low Reynolds number aerofoils with significantly different aerodynamic characteristics on a small MAV. These characteristics are taken from test data when available, and/or estimated by the XFOIL code of Drela \cite{Drela1989}. A total of 8 optimisation cases are studied for the purpose, leading to 6 different options for new wing planforms (and associated twist distributions along the wing span) with an improved performance.  It is found that the improvements in drag coefficient using the PROWING code are about 5\%.  However, by allowing the operating lift coefficient $C_L$ to float within a specified range, drag bucket characteristics of the Eppler E423 aerofoil used on Avion can be exploited to improve the endurance, which is a major performance parameter for Avion.  Thus, compared to the control wing $W_0$ (with operating point at $C_L$ =0.7) used in the preliminary design, permitting a variation of $C_L$ over a range of $\pm$ 10\% is shown to enhance the endurance of wing $W_4$ by 18.6\%, and of wing $W_{6}$ with a permitted $C_L$ range of $\pm$ 50\% by 39.2\%. Apart from the philosophy of seeking optimal operating conditions for a given configuration, the advantages of optimising design parameters such as washout of a simple wing proposed in the preliminary design stage, is also demonstrated.
\end{abstract}
	\maketitle
	\section{Introduction}
	A wing optimisation philosophy has been developed by Rakshith \emph{et al.} \cite{Rakshith2015} (referred to as RDNP from hereon) for tractor configuration turboprop aircraft. It exploits the local increase in velocity over the wing in the slip\textendash stream due to the propeller. It was found that if the slip\textendash stream is taken into account, the optimal wing for a given cost function (e.g. drag) at a given lift has novel spanwise distribution of chord and twist. The idea was to use the downwash created by the propeller to calculate how much wing downwash can be removed for a given lift to be produced. The optimisation was done through PROWING, a software developed for the purpose, which permits induced, total or viscous drag coefficient to be reduced by suitably changing the cost function that is minimised in the optimiser code.
	
	The CSIR National Aerospace Laboratories (NAL) at Bangalore, India have designed a tractor propeller micro aerial vehicle (MAV) called Avion. If the same philosophy as RDNP can be applied to the MAV, the benefits on flight performance can be investigated at much lower cost and effort, as compared to regional transport aircraft (RTA) of RDNP. But the MAV is much smaller and slower than the RTA; for example the flight Reynolds number and speed of Avion during cruise are only \num{294000} and 15 m/s respectively, whereas for the RTA the cruise Reynolds and Mach numbers are about $6 \times 10^6$ and 0.44 respectively. Some of the tools in the optimisation process of PROWING, therefore, could well be replaced for the MAV. For example, the numerical solver used for obtaining the propeller slip\textendash stream is changed to an incompressible one. 
	
	At the low Reynolds number of Avion the flow is laminar and the aerofoils are characterised by a low drag region (laminar drag bucket) over an appreciable range of angle of attack. Furthermore, for MAVs designed for surviellence operations an important performance parameter is endurance, which is proportional to the lift to drag ratio $L/D$, also referred to as endurance factor here. This is an additional parameter that will be optimised in this paper as it is linearly related to the range of a battery powered aircraft (\emph{cf.} \cite{ref15} and equation \eqref{Eq:Range_Eq} below).
	
	The rest of the paper is organised as follows. Section \ref{sec:Propeller} describes the methodology to obtain the propeller slip\textendash stream for this incompressible flow. Section \ref{sec:Avion_char} discusses the  characteristics of Avion in its original form as designed by NAL. The optimisation methodology and results are illustrated in section \ref{sec:Optimisation} and the conclusions are presented in section \ref{sec:Conclusions}.

\section{Propeller Performance Parameters}\label{sec:Propeller}
The aircraft wing is immersed in the flow field of the propeller, also referred to as propwash. The aerodynamic forces over the wing are directly influenced by propwash and this must be evaluated accurately. It can be obtained either computationally or experimentally. Any experiment that provides velocity vectors in the plane perpendicular to the streamwise direction (e.g. 2D PIV) is adequate for the purpose. 

On the computational side, RDNP used an in\textendash house code that solves Euler equations for compressible flow (termed as PROP\textendash EULER). This cannot be directly used here as the air flow at the low speeds (around 15 m/s) encountered by the MAV is essentially incompressible. Instead an implementation of an open source software, OpenFOAM is shown here, using a function for OpenFOAM called ``RotorDiscSource" developed for modelling flow around a helicopter \cite{Wahono2013}. It models the propeller as a geometrically simpler actuator disc of finite thickness. Each cell within this disc is a momentum source that injects energy into the flow. The actuator disc model requires propeller characteristics such as the aerofoil, the chord and twist distribution along the propeller blade, number of blades, and the rotation rate as input. This methodology is essentially similar to the PROP\textendash EULER code, but is specially designed for OpenFOAM. Aerofoil data consists of a look\textendash up table for the lift and drag characteristics at each span\textendash station. This can be taken from 2D aerofoil wind tunnel tests or a panel method such as XFOIL \cite{Drela1989}. Using OpenFOAM allows flexibility in terms of choosing the appropriate flow solver. Here, an incompressible Euler solver is adequate as turbulence is not of utmost importance (similar to PROWING). This particular procedure is validated against two propellers for which experimental data are available in the public domain. This is described next.
\subsection{Validation of computational procedure proposed}
Two propellers of very different size and application are chosen to demonstrate the robustness of the present procedure. The first is a propeller similar in size to that used on the AVION, and the second is from RDNP for the much larger RTA.
\subsubsection{Deters' propeller\textendash DA4002}
Deters \textit{et al.} \cite{Deters2014} report some experiments on commercially available small propellers with sizes varying from $2.25$ to $9\  \text{in.}$ in diameter. Aerofoil profile shapes of off\textendash the\textendash shelf propellers are hard to obtain, but they are generally optimised for efficiency. Deters \textit{et al.} also designed some propellers for which the aerofoil shape was known. One of these propellers is chosen here as a validation case. The lift and drag polars for the aerofoils are generated via XFOIL \cite{Drela1989}. The aerofoil is referred to as DA4002 (9x6.75) in \cite{Deters2014}, the notation indicating propeller diameter and pitch (distance moved forwards in one revolution at the design point) of $9\ \text{in.}$ and $6.75\ \text{in.}$ respectively.

The propeller uses SD1075 aerofoil for the most part, changing to SD1100 near the tip. However, as the blending function near the tip is not provided in \cite{Deters2014}, SD1075 is assumed here throughout the propeller. The propeller has 2 blades, each of constant chord, and a twist decreasing smoothly from about $42^\circ$ near the root to about $14^\circ$ near the tip (further details can be found in \cite{Deters2014}). The propeller is simulated for various values of the advance ratio
\begin{equation}
J=\frac{V_\infty}{nD},
\end{equation}
where $V_\infty$ is free\textendash stream velocity, $n$ is revolutions per second and $D$ is propeller diameter. $J$ is varied by changing $V_\infty$, while keeping $n$ at 83.3 r.p.s. (5000 r.p.m.). For this propeller $V_\infty$ is simulated over the range $3.8\ \text{m/s}$ $(J=0.2)$ to $17.15$ m/s $(J=0.9)$. 

\begin{figure}[h!]
	\centering
	\includegraphics[width=0.49\textwidth]{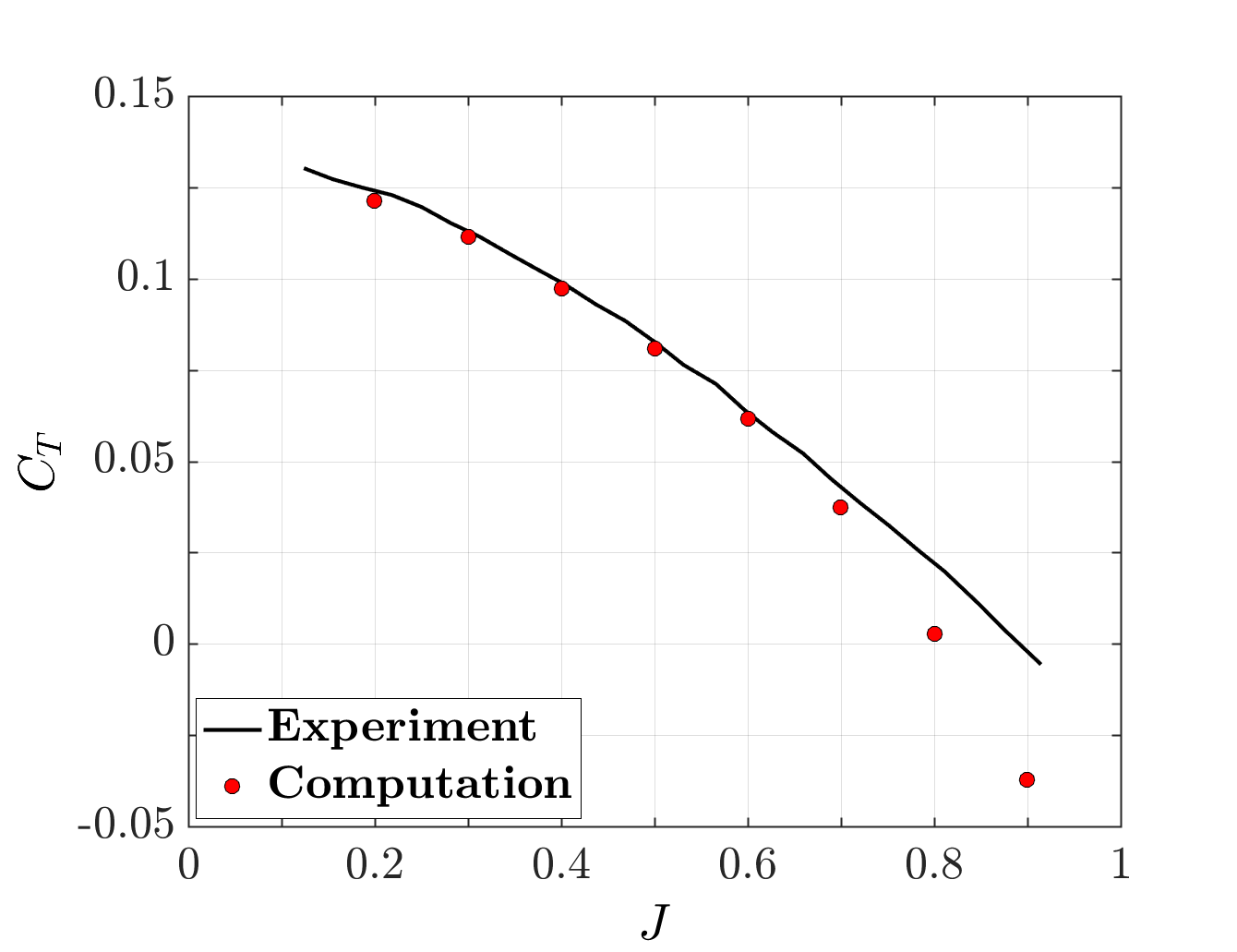}
	\caption {Experimental \cite{Deters2014} and computational (OpenFoam Incompressible Euler) thrust coefficient for DA4002}
	\label{fig:F2}
\end{figure}

Figure \ref{fig:F2} shows the comparison between the experimental \cite{Deters2014} and computational values of the thrust coefficient
\begin{equation}
C_T=\frac{T}{\rho n^2 D^4},
\end{equation}
of the Deters propeller, where $T= \text{thrust}$ and $\rho= \text{density}$. 
The computational and experimental results match well for advance ratios up to $0.7$. The small difference in the values at higher values of $J$ can be partly explained by the bending of the propeller blades in the experiments \cite{Deters2014}, which changes the effective angle of attack on the aerofoil. As the free\textendash stream velocity is increased at constant propeller r.p.m. the effective angle of attack on the blade decreases, and at the highest speeds some of the blade sections near the tip will have more negative angles of attack for the rigid blade (as assumed in the computations) relative to the actually more flexible blade used in the experiment. Furthermore, the lift and drag polars for the current simulation are obtained from XFOIL, which does not model the abrupt stall observed on the pressure side very well at the very low Reynolds numbers of interest here ($2 \times 10^4$ to $9 \times 10^4$) \cite{Aerofoil_summary}. Hence, a stall could begin to appear on the pressure side of the aerofoils in the computations. These two phenomena explain the sudden drop in thrust at $J>0.7$. Another possible source of error is the approximate modelling of the aerofoil at the bent tip mentioned above. However, the design point for this or any other propeller is generally away from the conditions where any of the blade sections are near stalling angles. Hence, the above method should be sufficient for normal operating conditions. But stall characteristics may be important at take\textendash off and landing.
\subsubsection{NACA Propeller}
\begin{figure}[h!]
	\centering
	\includegraphics[width=0.49\textwidth]{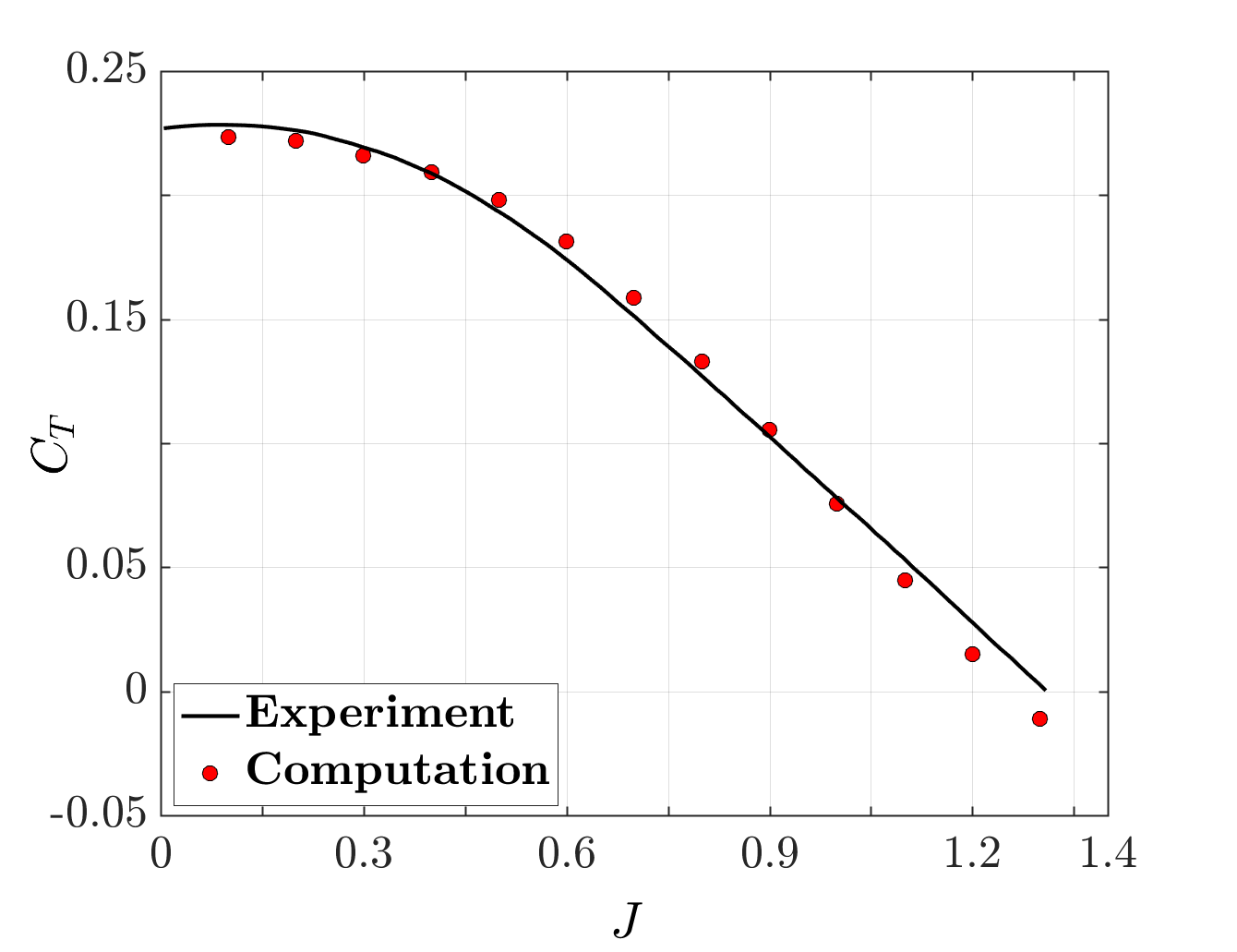}
	\caption {Experimental \cite{Hartman1938} and computational (OpenFoam Incompressible Euler) thrust coefficient for NACA propeller}
	\label{fig:F5}
\end{figure}

This is the same 4\textendash bladed propeller as used by RDNP. Geometric and experimental data for this variable pitch NACA propeller (as it will be called here) are available in \cite{Hartman1938}. The blade is $10$ ft. long and has blade sections that are RAF 6 aerofoils of varying thickness to chord ratio. A propeller with a pitch setting of $25^\circ$ at $75\%$ blade radius was considered for the validation run. A propeller rotation speed of 1000 r.p.m. was chosen, varying the advance ratio from $0.1$ to $1.3$ as the free\textendash stream velocity changes from  $5.08$ m/s to $66.04$ m/s. Figure \ref{fig:F5} compares the experimental \cite{Hartman1938} and computational results for the thrust coefficient, and shows good agreement between the two.
For the speeds considered, the Reynolds number (based on local chord and the vector sum of angular and free\textendash stream velocity) varies from \num{63000} to \num{480000} over the blade. However, the lift and drag polars adopted from the work of RDNP here were calculated for $J=1.0$ only. This can explain the mild crossover between the experimental and the computational curves at $J=1.0$. Another posible source of error is the slight difference in actual and specified pitch setting that can occur in a variable pitch propeller. Flow solver used is incompressible Euler, hence compressibility effects can add to the errors as the tangential speed at the tip for the chosen r.p.m. is $319$ m/s and the highest free\textendash stream velocity simulated is $66$ m/s. Therefore, at the rightmost data point on figure \ref{fig:F5} the propeller tip experiences an incident speed of $326$ m/s.

The code used here has the capability to provide slip\textendash stream information over a prescribed volume around the propeller, but it relies on the availablity of the propeller geometry, i.e. aerofoil shape, twist and chord distribution. In case these are not available, experimental methods must be used. This is indeed the case for the Avion propeller as the required data are not available from the propeller manufacturer.

\subsection{Experimental Data: 2D PIV}
 An APC 11$\times$8 thin electric propeller was proposed for the Avion project. But due to its unavailability at the time of design, the very similar APC 11$\times$7 was used. Both are 11 in. in diameter, the only difference being that the pitch is 8 in. for the former and 7 in. for the latter. From the extensive test data on these propellers \cite{Propeller_Database} it is seen that the dynamic (non\textendash static) thrust for the two propellers is very similar over the range of advance ratios from $J\approx 0.1$ to 0.8. Hence the slip streams of interest should not be very different from each other. 

\begin{figure}[h!]
	\centering
	\includegraphics[width=0.49\textwidth]{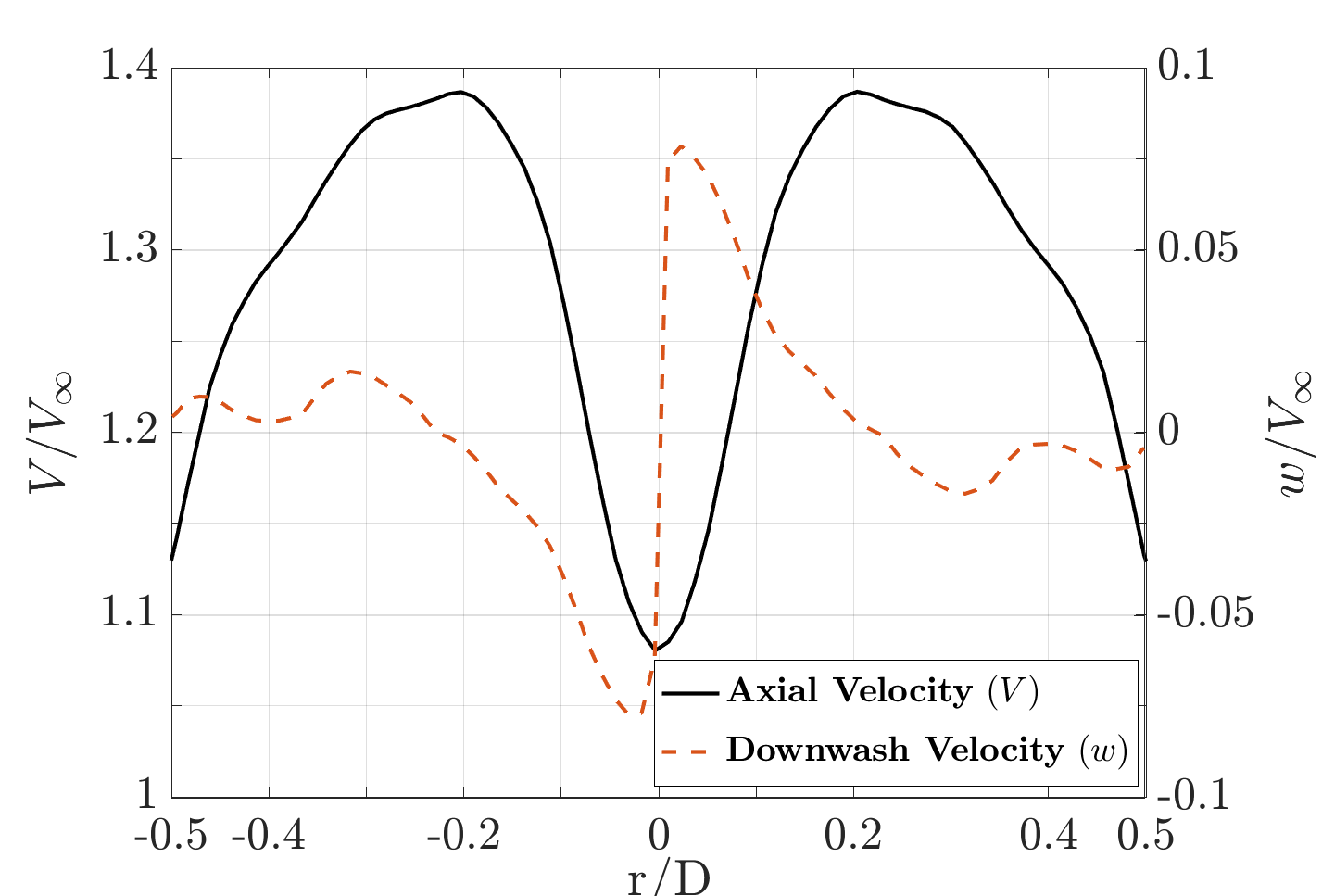}
	\caption {APC 11$\times$7 thin electric propeller slip\textendash stream velocity one diameter downstream of the propeller plane.}
	\label{fig:F10}
\end{figure}

A 2D PIV experiment has been performed at CSIR\textendash NAL \cite{NAL_Personal} on the APC 11$\times$7 propeller where the axial and downwash velocities are captured along a vertical line in each of the propeller half planes, respectively  at $0.571$, $0.823$, $1.075$ and $1.3239$ diameters downstream of the propeller. In the experiment, an upstream blockage affected the flow in the lower half plane downstream of the propeller. Therefore, the velocities from the upper plane are mirrored to obtain the velocity vectors along the vertical line at each plane. An interpolation of each of the velocities (axial and downwash) can be performed to obtain the slip\textendash stream over a 2D plane in case a higher fidelity panel method is used to calculate the aerodynamic forces over the aircraft wing. As the  lifting line theory is used to evaluate the aerodynamic forces in the current exercise, the spanwise distribution of the velocities at a particular downstream station (corresponding to the quarter chord of the wing) is sufficient. For simplicity, the propwash at one propeller diameter  downstream of the propeller (as illustrated in figure \ref{fig:F10}) is used. This corresponds to a location less than 65\% of the local wing chord length at the propeller span\textendash station (30\% along the wing semi\textendash span from the root). The precise location cannot be quoted as the position of the propeller relative to the wing leading edge in the flow direction is unknown.

\section{Avion control characteristics}\label{sec:Avion_char}
\begin{figure}[h!]
	\centering
	\includegraphics[width=0.49\textwidth]{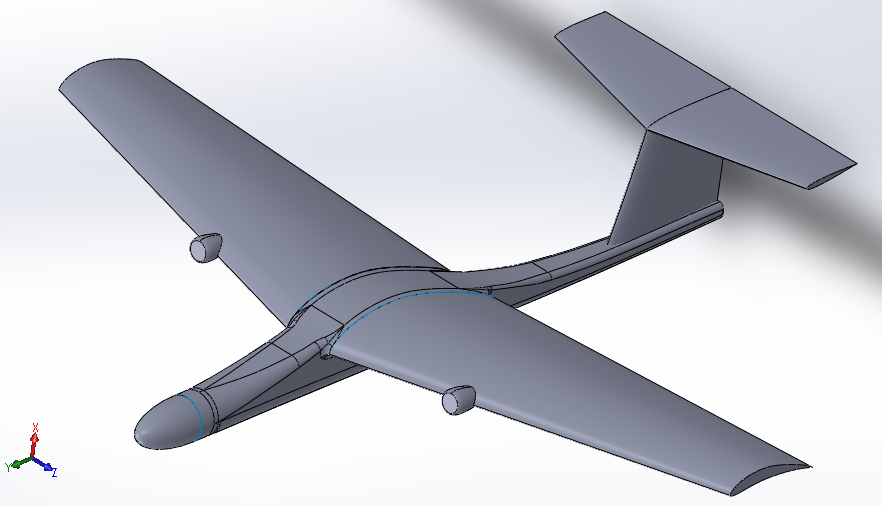}
	\caption {Initial Configuration of Avion \cite{NAL_Personal}}
	\label{fig:F1_4}
\end{figure}

Avion is a very good case to test the philosophy of optimising wing design by exploiting the propeller slip\textendash stream, studied by RDNP. This is because wind tunnel tests are feasible for the MAV on flight scales. Figure \ref{fig:F1_4} illustrates the standard control configuration of Avion and table \ref{table:Table_1} lists the characteristics of a prelimenary design of the Avion wing at NAL.

\begin{table}[h!]
	\centering 
		\renewcommand{\arraystretch}{1.4}	
	\begin{tabular}{C{2.5cm}|C{1.5cm}||C{2.5cm}|C{1.5cm}} 
		\bfseries{Wing parameter} &\bfseries{Avion}& \bfseries{Operating parameter} & \bfseries{Avion} \\  \hline 
		Aerofoil & Eppler E423 &Wing Loading ($W$) & 74.39 N/m$^2$  \\ 
		Aspect Ratio ($AR$)&5.35&Stall Speed ($V_\text{stall}$)&8.4 m/s @  $C_L$=2.0\\
		Area ($S$) & 0.479 m$^2$ &Cruise Speed ($V_{\infty}$)&15 m/s @ $C_L$=0.7 \\		
		Span ($s$) & 1.60 m &Design Altitude &1000 m\ ASL\\		
		LE Sweep ($\Lambda$) & 14$^\circ$ &Cruise Reynolds no. based on m.a.c. & \num{294000}\\		 
		Dihedral and Washout ($\gamma$) &0$^\circ$&& \\
		Mean aero. chord (MAC)&0.310 m& &\\
	\end{tabular}
	\caption{Characteristics of a preliminary design of Avion with mass 3.64 kg and propeller positioned at 30\% along the wing span from wing root. Propeller is assumed to be rotating upwards in\textendash board such that it creates an additional upwash in\textendash board and downwash out\textendash board \cite{NAL_Personal}.\label{table:Table_1}} 
\end{table}  
The most important features that must be evaluated before deciding upon a tool to calculate the aerodynamic forces generated by Avion are the aerofoil and the wing characteristics. These are discussed next.

\subsection{Aerofoil Characteristics}\label{sec:XfoIL}
\begin{figure}[h!]
	\centering
\includegraphics[width=0.4\textwidth]{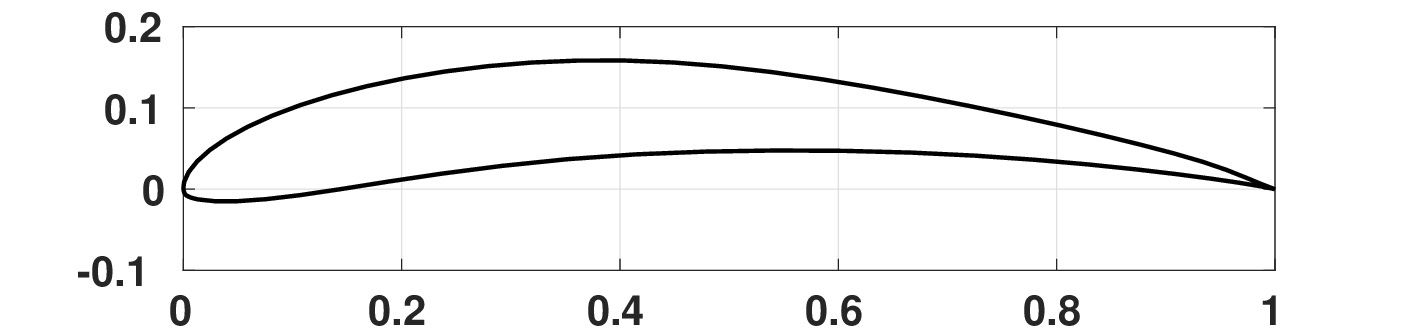} \\
	\caption {E423 aerofoil geometry} \label{fig:e423shape}
\end{figure}

Avion uses the Eppler E423 aerofoil (figure \ref{fig:e423shape}) throughout its span, therefore it is essential to evaluate its lift and drag characteristics accurately. Lift and drag forces generated by the aerofoil are functions of the angle of attack and can be obtained either from wind tunnel tests or from a computational tool such as XFOIL \cite{Drela1989}. In the current optimisation exercise (section \ref{sec:Optimisation}) experimental data from \cite{Aerofoil_summary} is used.

\begin{figure}[h!]
	\centering
	\subfloat[Lift curve]{\includegraphics[width=0.49\textwidth]{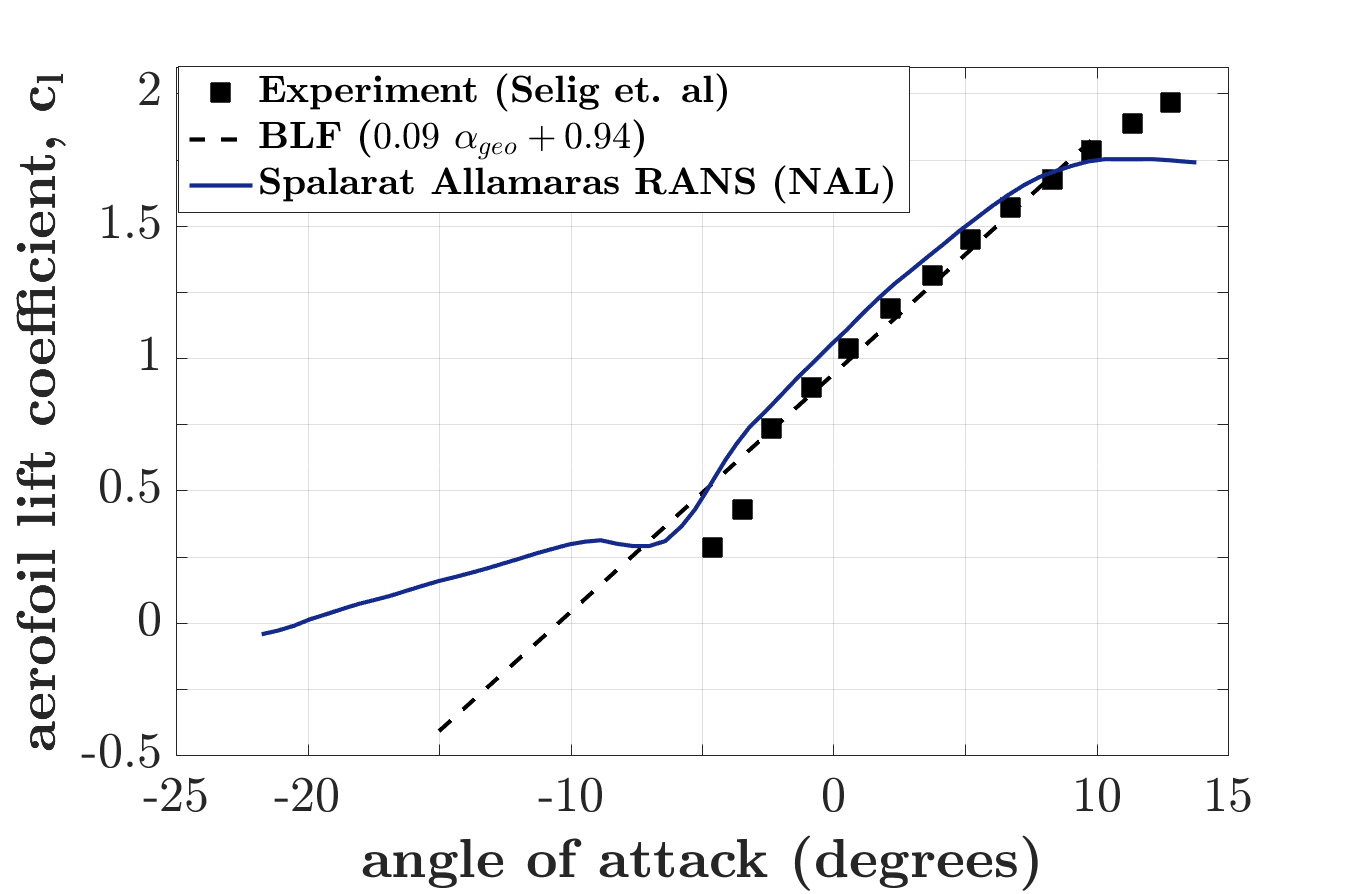}\label{fig:F15}} \\
	\subfloat[Drag curve]{\includegraphics[width=0.49\textwidth]{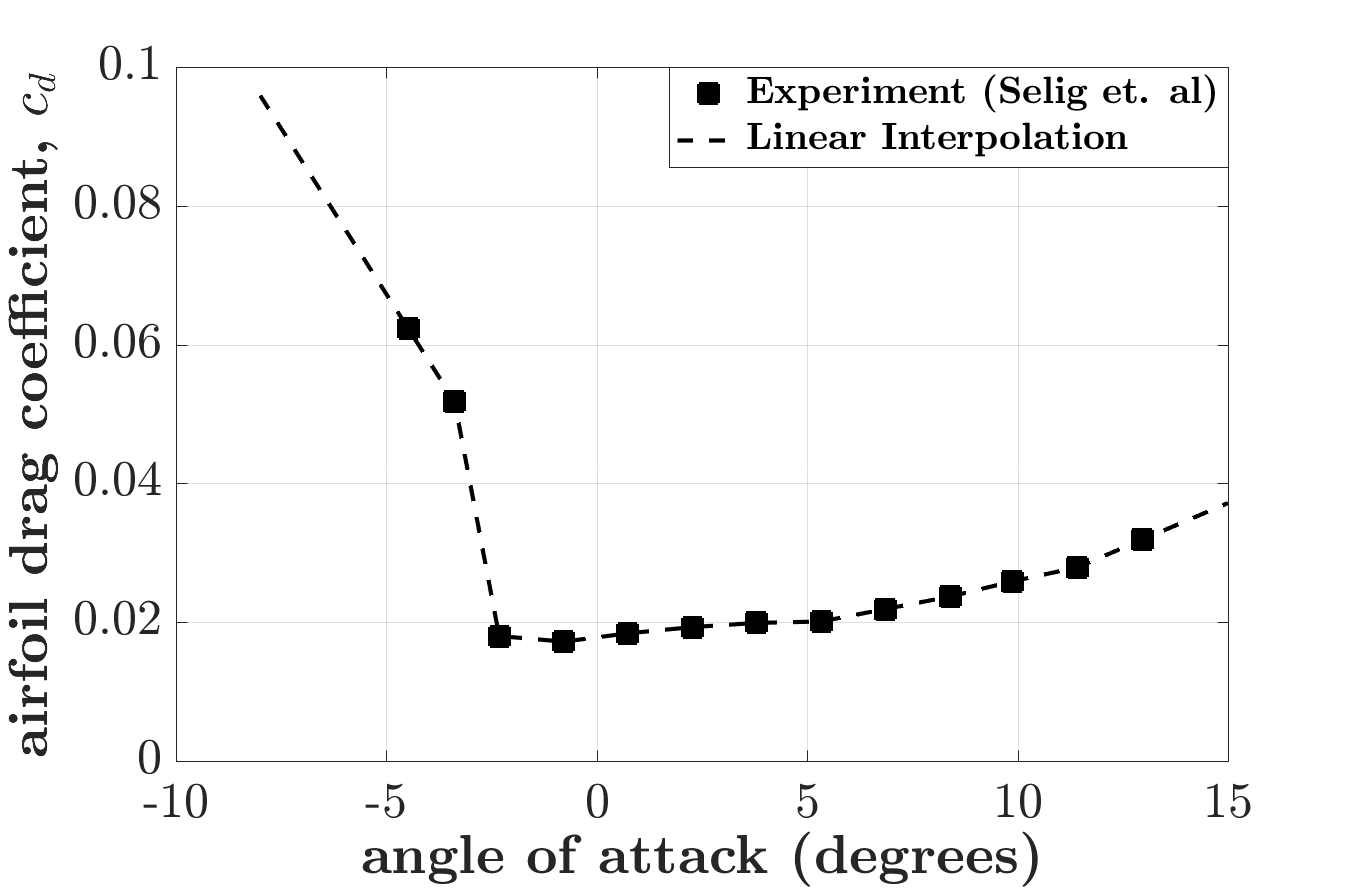}\label{fig:F12_1}} 
	\caption {E423 aerofoil characteristics, $Re=\num{300000}$ \label{fig:F12}}
\end{figure}

The E423 is designed to provide high maximum lift, and the high effective camber due to the displacement thickness aids in achieving this \cite{Aerofoil_summary}. Experimental 2D lift and drag characteristics of E423 were recorded by Selig \textit{et al.} \cite{Aerofoil_summary} and are shown here in figure \ref{fig:F12}. The lift curve based on Reynolds Averaged Navier Stokes equations (RANS) using the Spalart\textendash Allamaras model for turbulence viscosity calculated at NAL \cite{NAL_Personal} is also shown for reference in figure \ref{fig:F15}. From figure \ref{fig:F12_1} it can be observed that the drag on the E423 aerofoil remains low for a large range of about 15$^\circ$ in angle of attack. This characteristic will be exploited in the optimisation methodology in Section \ref{sec:Optimisation}.
\subsection{Wing Characteristics}\label{sec:wing}
\begin{figure}[h!]
	\centering
	\includegraphics[width=0.49\textwidth]{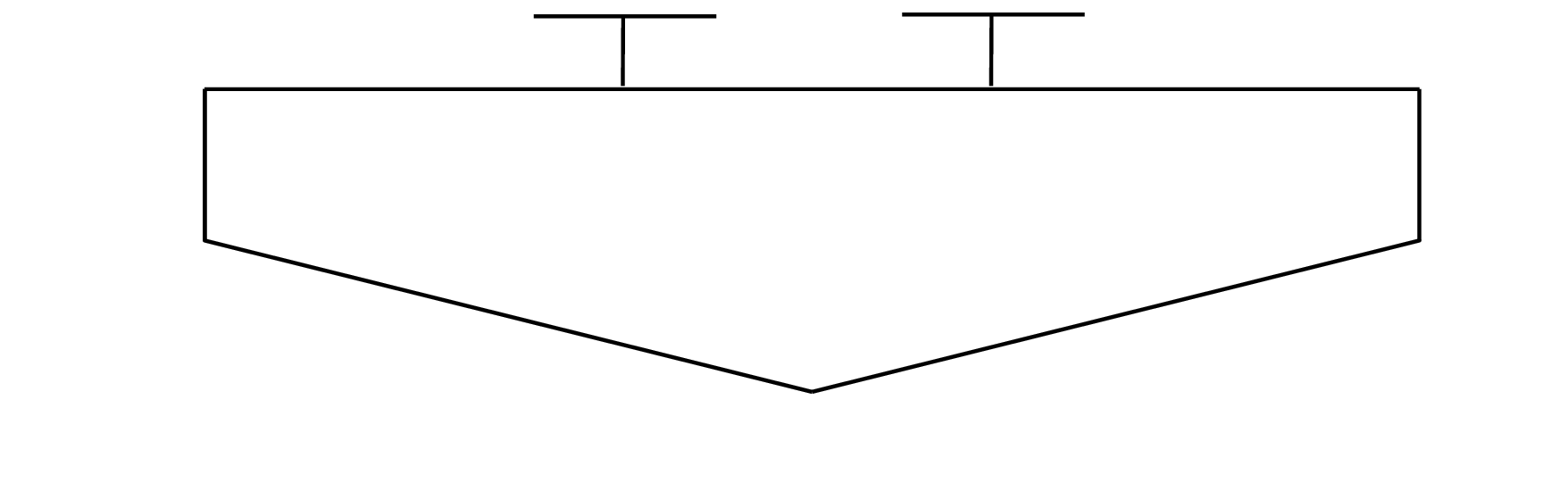}
	\caption {Schematic of Original Avion Wing, $W_0$\label{fig:Orig_wing}}
\end{figure}

Following RDNP, the aerodynamic loads over the complete wing are calculated using the lifting line theory (LLT) modified to include the propwash. 
Avion has a leading edge (LE) sweep of $14^\circ$  that was incorporated in the original design for the sole purpose of having a trailing edge perpendicular to the fuselage (for easier accommodation of the control devices in the final design \cite{NAL_Personal}). But, as LLT assumes no sweep (strictly no 1/4 chord sweep), the wing profiles drawn in this section have no LE sweep and the chord modifications are applied to the trailing edge. The schematic of the original wing $W_0$, shown in figure \ref{fig:Orig_wing}, serves as a reference planform for the optimised geometries (section \ref{sec:Optimisation}). Thus, dimensions and the other geometric properties such as the angle of attack and leading edge sweep are the same as listed in Table \ref{table:Table_1}. The flow over the wing in figure \ref{fig:Orig_wing} is from the top of the page. 
Assuming that the aerofoil operates in a linear $C_L$\textendash $\alpha$ range that extends over a sufficiently wide range of angles of attack, the sectional/ aerofoil lift coefficient can be expressed as
\begin{equation} \label{eq:Lift_Eq}
c_l(y)=a_\text{0}(y) (\alpha_\text{eff}(y)-\alpha_\text{0}(y) ),
\end{equation}
where $a_\text{0}(y)$, $\alpha_\text{eff}(y)$ and $\alpha_\text{0}(y)$ are the lift curve slope, effective angle of attack and the zero lift angle of the aerofoil respectively at the spanwise location $y$. Angle $\alpha_\text{eff}(y)$ is given by
\begin{equation} \label{eq:Effect. Angle}
\alpha_\text{eff}(y)=\alpha_\text{geo}+\alpha_\text{twist}(y)-\alpha_\text{downwash}(y)-\alpha_\text{prop}(y),
\end{equation}
where $\alpha_\text{geo}$ is the geometric angle of attack of the wing, and $\alpha_\text{twist}(y)$, $\alpha_\text{downwash}(y)$ and $\alpha_\text{prop}(y)$ are the sectional twist, wing downwash and propwash angles respectively at the spanwise location $y$.

The aerofoil $c_l$ vs. angle of attack ($\alpha_\text{eff}(y)$) curve is not linear at angles lower than about $-2.5^\circ$ as shown in figure \ref{fig:F15}. This coincides with the most negative angle of attack of the laminar drag bucket region in figure \ref{fig:F12_1}. However, at angles of attack larger than $-2.5^\circ$, the experimental lift curve in figure \ref{fig:F15} is well approximated with the best linear fit (BLF) line. This linear function for the sectional lift curve is used in the LLT calculations, with a conservative choice of 320 collocation points and 48 Fourier modes. RDNP demonstrated convergence at much smaller values for the optimised wings and this was also checked here for the results of Section \ref{sec:Optimisation}.
\begin{figure}[h!]
	\centering
	\subfloat[Lift Curve]{\includegraphics[width=0.49\textwidth]{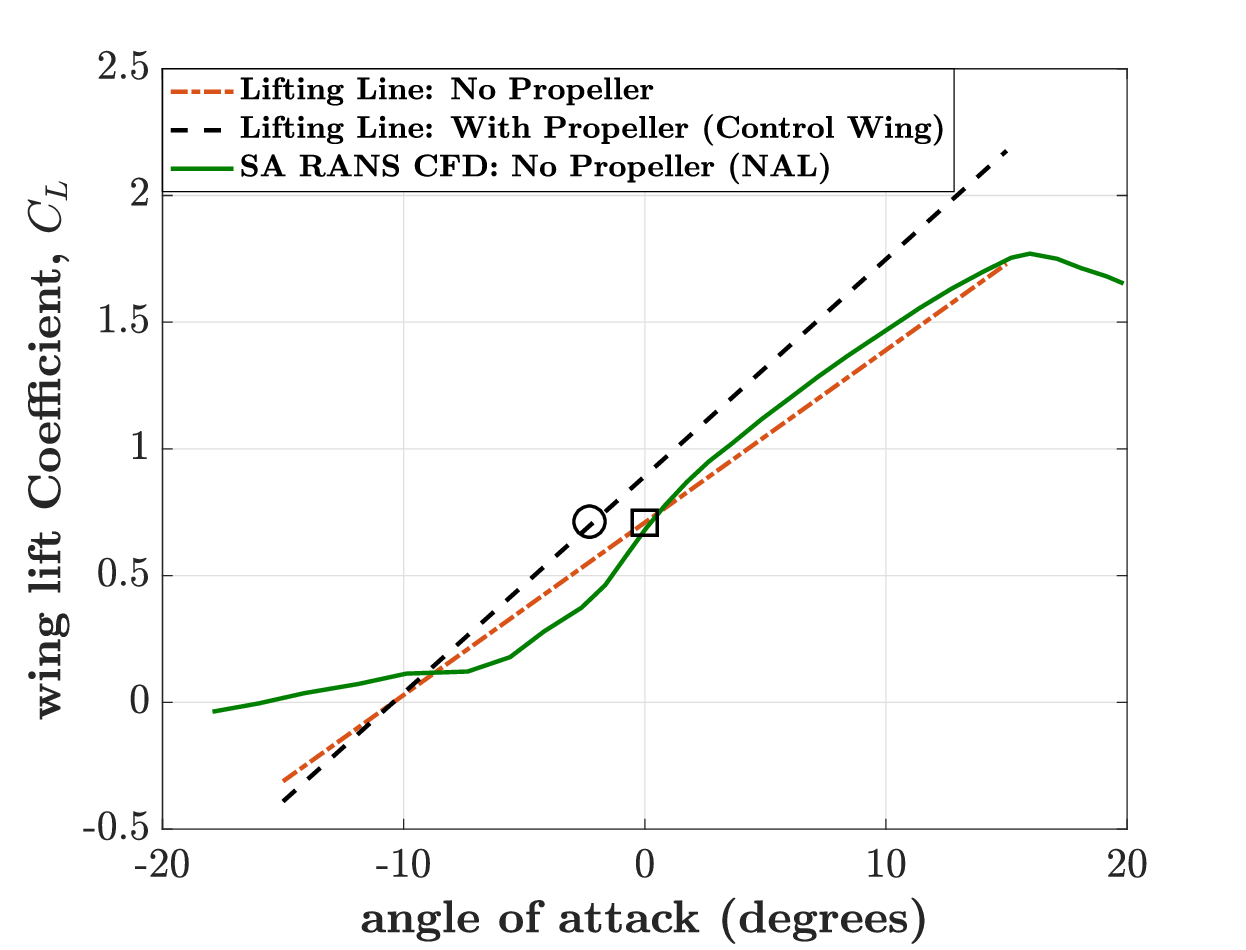}\label{fig:F16}} \\
	\subfloat[Drag Polar]{\includegraphics[width=0.49\textwidth]{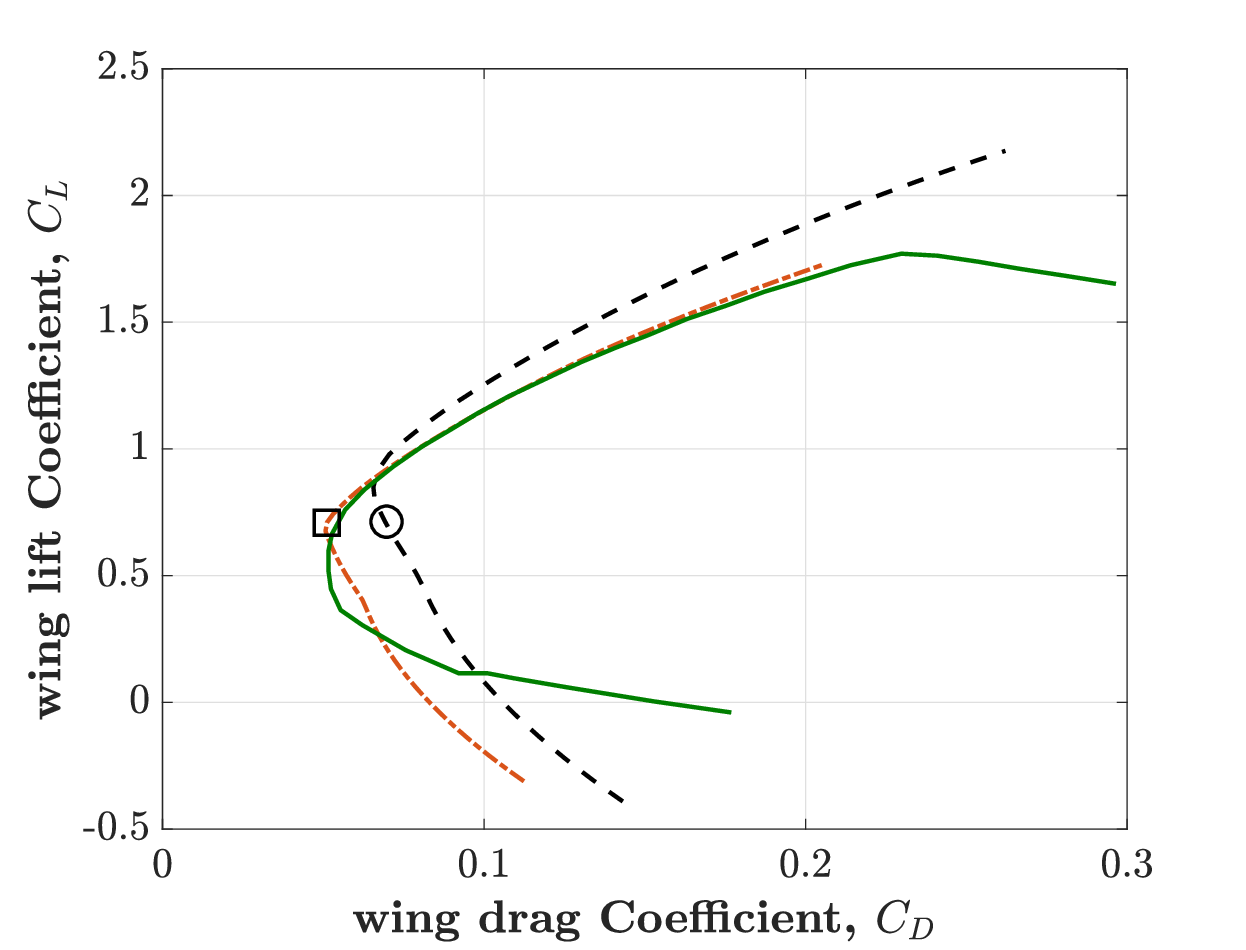}		\label{fig:F17}}
	\caption {Avion wing characteristics. The legends for the curves are the same for both plots: \Square\hspace{0.02in} and \Circle\hspace{0.02in} represent the operating $C_L=0.7$ for the wing without and with propeller effect respectively.\label{fig:WingPolars}}
\end{figure}

The Avion wing is designed to cruise at aircraft lift coefficient $C_L=0.7$ at $\alpha=0^\circ$, but the propeller slip stream was not taken into account in the design \cite{NAL_Personal}. Upon incorporating propeller effects (propwash), the lift curve of the wing shifts upwards as observed from the LLT curves in figure \ref{fig:F16}. This is due to higher streamwise velocity in the slip stream aided by axial velocity of the propeller. The change in slope is perhaps due to the downwash velocity of the propeller that aids stall at some of the spanwise sections experiencing lower angles of attack just out\textendash board of the propeller and near the wing tip. In figure \ref{fig:F16} the non\textendash linear effects captured by the CFD simulation of the whole wing using Spalart\textendash Allamaras RANS model (SA RANS CFD) at NAL are ignored in the LLT calculation. Increased lift taking account of the propeller\textendash effect means that the geomteric angle of attack of the wing, $\alpha_\text{geo}$ has to be reduced to achieve the design $C_L$. This could be dangerous as the aerofoil experimental results in figure \ref{fig:F15}, and the SA RANS CFD results in figure \ref{fig:F16}, show that lift falls rapidly at negative angle of attack, possibly due to separation on the pressure surface. As expected this is accompanied with an increase in drag coefficient in figure \ref{fig:F17}. 

In the lifting line model only the viscous and induced drag are incorporated. Viscous/ profile drag over the whole wing is evaluated by taking the sum of drag acting on the individual aerofoil sections:
\begin{equation} \label{Eq:DragEquations}
D_P=\rho\int_{-s}^{s}\frac{1}{2} V(y)^2c_d(y)c(y)\ dy,
\end{equation}
where $c_d(y)$ is the profile drag coefficient for the aerofoil section at spanwise location $y$. It is obtained from a $c_l-c_d$ interpolation of the data in figure \ref{fig:F12}.

Wing lift and the drag polars for the no\textendash propeller case from RANS and LLT match well for $C_L$ greater than $0.7$ or positive angles of attack (figure \ref{fig:WingPolars}). The significant departures at negative angles of attack could be either due to the non\textendash linear effects outside the realm of the LLT, or the effect of small scale features over the wing that RANS is not able to fully resolve, but are incorporated into the LLT through the experimental data in figure \ref{fig:F12}. The extrapolation of the polar to negative angles of attack in figure \ref{fig:F12_1} is not expected to play a large role, at least at the onset of this discrepancy at the wing angle of attack $\alpha_\text{geo}$ of $0^\circ$, as BLF for the $c_l$\textendash $\alpha$ curve closely follows the RANS results for angle of attack larger than $-7.5^\circ$ in figure \ref{fig:F15}. Also, even after considering $\alpha_\text{downwash}$ and $\alpha_\text{prop}$,  $\alpha_\text{eff}$ for most of the aerofoil sections would be more positive than the last data point available at $-5^\circ$ in figure \ref{fig:F12_1}. However, at the even lower wing angles of attack where LLT predicts less $C_D$ for same $C_L$ in figure \ref{fig:F17}, extrapolation of drag at negative angles in figure \ref{fig:F12_1} and the BLF line in figure \ref{fig:F15} may need to be reconsidered. However for the current exercise this is not attempted.

The LLT curves in figure \ref{fig:F17} reveal that the addition of the propeller leads to a reduction in $C_D$ for any given $C_L$ larger than 0.85. This improvement arises in both viscous and induced drag coefficient. Viscous drag coefficient improves as $\alpha_\text{eff}$ on more of the aerofoil sections along the span is in the low drag region in figure \ref{fig:F12_1}. Induced drag coefficient improves because the same lift can be produced at a lower angle of attack once the propeller slip\textendash stream is accounted for (figure \ref{fig:F16}). $C_D$ increases at negative wing angles of attack $\alpha \le -2.5 ^\circ$,  because the additional downwash of the propeller creates a large (negative) enough $\alpha_\text{eff}(y)$ for the aerofoil sections immediately out\textendash board of the propeller, such that in terms of the aerofoil drag ($c_d$) curve these sections lie left of the low drag region in figure \ref{fig:F12_1}.

\section{Optimisation Methodology and Results}\label{sec:Optimisation}
The optimisation philosophy in the current exercise is similar to that of RDNP, however certain features have been modified to exploit the characteristics of AVION. 

The optimisation of RDNP was based on fixed aerodynamic constraints such as wing lift and pitching moment, along with other geometric constraints such as tip and root chord, wing area, span and bounds on chord and twist to ensure a manufacturable wing. Structural constraints (such as root bending moment) could also be specified separately if necessary. The optimiser varies the chord and twist distributions (based on Bezier parameterisation in  RNDP), i.e. the control parameters that minimise a cost function (which could be either total or induced drag coefficient). However other choices for control parameter constraints and the cost function are possible.

In order to get smooth and manufacturable wing profiles 4 Bezier modes for both the chord and the twist distribution are used. The constraints on the wing geometry that remain same throughout are the wing area and the chord length at root and tip of the initial/ control wing. Two approaches for optimisation of the original control wing specified in table \ref{table:Table_1} are documented. The first approach is the same as that of RDNP i.e. wing shape is altered whilst keeping $C_L$ constant at the original design level. However, in the second approach $C_L$ is allowed to vary. Cases with a 10\% and a 50\% change in $C_L$ relative to the original design are studied for the possible optimisation gains. For the 10\% case the effect on operating conditions is relatively insignificant. 

Following the optimisations from the aforementioned two approaches certain changes to the original design specifications in table \ref{table:Table_1} are proposed and further optimisations are carried out on these.
\subsection{Approach 1: Fixed Wing Lift coefficient ($C_L=0.7$)}
For Avion at $C_L=0.7$ with propeller effect included, induced (case 1) and total (case 2) drag coefficient are optimised using the aerofoil characteristics shown in figure \ref{fig:F12}, obtained from experimental data. To obtain the operating $C_L=0.7$ specified in the initial design (table \ref{table:Table_1}), the wing geometric angle of attack is changed to $-2.25^\circ$. From hereon the change in drag coefficient is given by 
\begin{equation}\label{eq:drag_change_eq} \Delta C_D=C_{D,\text{NAL}}-C_{D,\text{opt}},\end{equation} 
where $C_{D,\text{NAL}}$ and $C_{D,\text{opt}}$ are the drag coefficient of the original and optimised wings respectively. Therefore, a positive $\Delta C_D$ implies a drag reduction. Results are summarised in Table \ref{table:Table_2}. Hereon, $C_{D_i}$, $C_f$ and $C_D(=C_{D_i}+C_f)$ refer to the induced, viscous and total drag coefficient of the wing respectively.
\begin{table}[h!]
	\centering 
	\renewcommand{\arraystretch}{1.4}	
	\begin{tabular}{C{3cm}||C{2.5cm}|C{2.5cm}} 
		\bfseries{Drag Parameter} &\bfseries{Case 1- $C_{D_i}$ Optimisation (figure \ref{fig:F19})}& \bfseries{Case 2- $C_D$ Optimisation (figure \ref{fig:F20})} \\  \hline 	
				Original $C_{D_i}$&\multicolumn{2}{c}{\num{0.0238}}\\	
		Original $C_{f}$ &\multicolumn{2}{c}{\num{0.0459}}\\	
		Original $C_{D}$ &\multicolumn{2}{c}{\num{0.0698}}\\	
		$\Delta  C_{D_i}(counts/\%)$&2.8/1.2\%&1.5/0.6\%\\	
		$\Delta C_{f}(counts/\%)$&17.3/3.8\%&37.1/8.1\%\\
		$\Delta C_{D}(counts/\%)$&20.0/2.9\%&38.6/5.5\%\\	
	\end{tabular}
	\caption{Avion optimisations for fixed $C_L$; 1 drag count= \num{0.0001}$C_D$\label{table:Table_2}} 
\end{table}
\begin{figure}[h!]
	\centering 
	\subfloat[$W_1$: Case 1 ($C_{D_{i}}$ optimisation)- $\Delta C_{D_i}= 1.2\%,\  \Delta C_f=3.8\%,\ \Delta C_D=2.9\%$]{\includegraphics[width=0.45\textwidth]{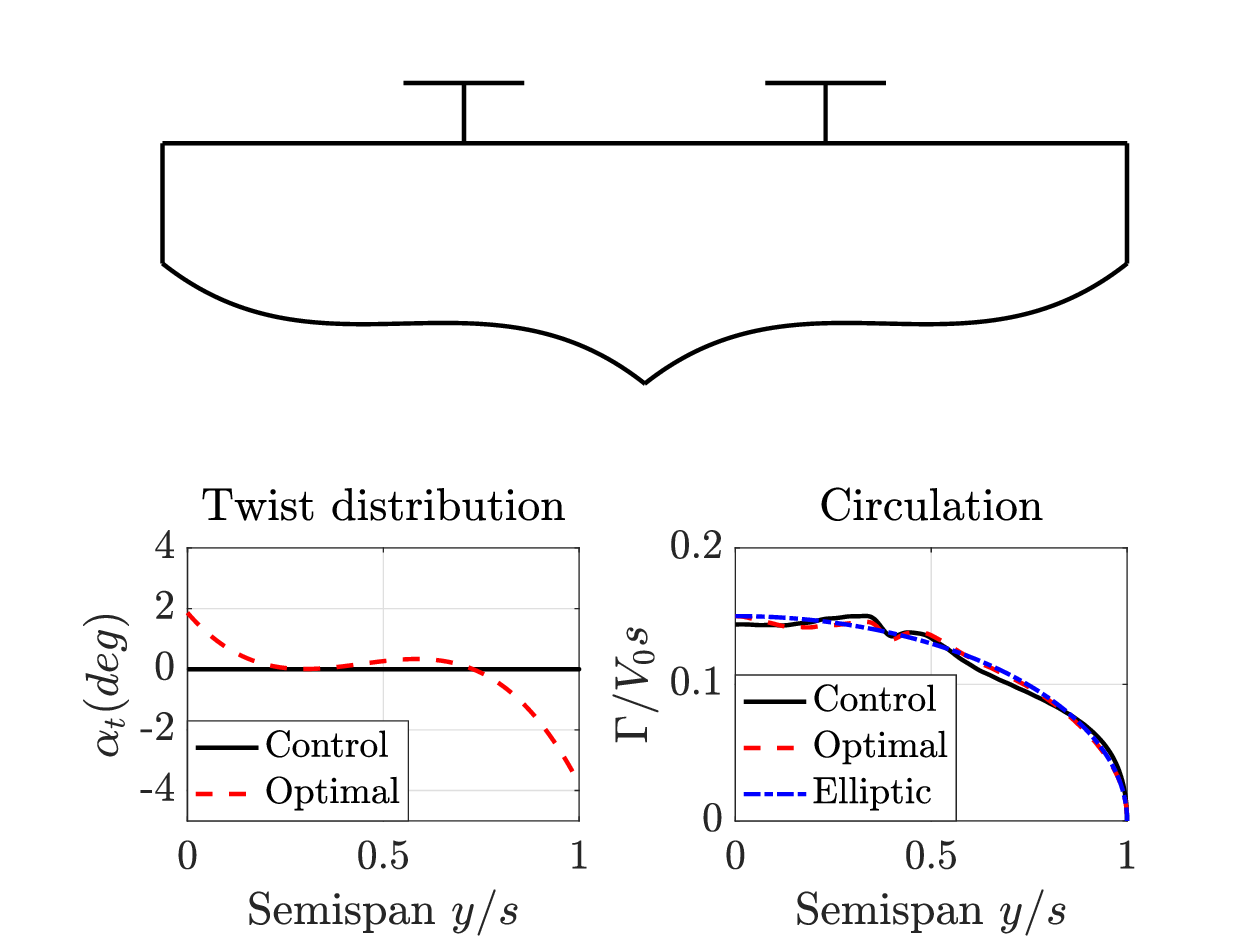}
		\label{fig:F19}}\\
	\centering    		
	\subfloat[$W_2$: Case 2 ($C_{D}$ optimisation)- $\Delta C_{D_i}= 0.6\%,\ \Delta C_f=8.1\%,\ \Delta C_D=5.5\%$]{\includegraphics[width=0.45\textwidth]{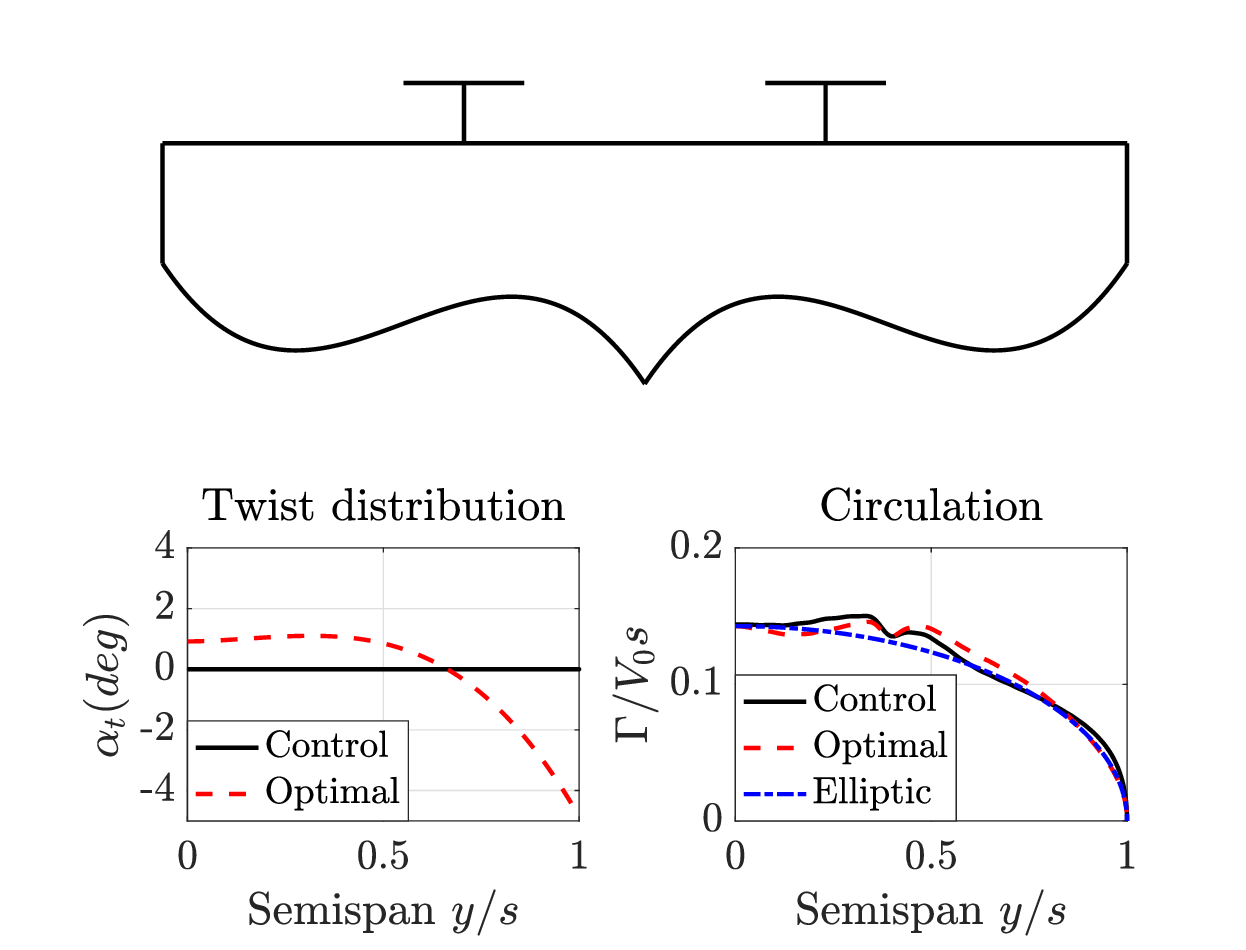} \label{fig:F20}}
	\caption {Planforms optimised with constant lift coefficient, $C_L=0.7$  @ $\alpha_\text{geo}=-2.25^\circ$\label{Fixed_CL_opt}}
\end{figure}
The observations for the $C_{D_{i}}$ and $C_{D}$ optimisation are the following.
\paragraph{\underline{$C_{D_{i}}$ Optimisation}:} The modified twist and chord distributions for optimal $C_{D_i}$ are shown in figure \ref{fig:F19} for wing $W_1$: the benefit in $C_{D_i}$  is only 1.2\%. Upon inspecting the circulation plot in figure \ref{fig:F19}, it can be noted that the shape of the circulation distribution of the control wing in the region outside the propeller span\textendash station is already quite close to the optimal elliptic load, which is the profile for minimum $C_{D_{i}}$ for wing mounted propeller wings as observed in RDNP. Hence, there is little room for improvement on this aspect. However, optimising for $C_{D_i}$ leads to a much higher (3.8\%) improvement in $C_{f}$. This is mainly due to the change in twist distribution in figure \ref{fig:F19}. The twist is increased at the wing root and also in the region just out\textendash board of the propeller, thereby bringing the aerofoil sections to lower drag regions of figure \ref{fig:F12_1}. The $C_{D}$ reduces by $2.9\%$.
	
Comparing with RDNP, it can be observed that improvement in terms of $C_{D_{i}}$ counts is comparable, in fact larger. In case 1 of RDNP, optimising for  $C_{D_i}$ led to a reduction in $C_{D_{i}}$ by $0.43$ counts ($1$ count $= 10^{-4}$), which in that case translated to a much higher $9.6\%$ reduction in $C_{D_i}$ because the aspect ratio of the wing was higher and the $C_{D_{i}}$ was only $\num{0.000407}$. We recall that according to the lifting line theory for untwisted wings on aircraft such as Spitfire with elliptic chord and circulation distribution along the span, $C_{D_{i}}$ is related to the $C_L$ and aspect ratio ($AR$) by
\begin{equation}\label{eq:cdi}
C_{D_{i}}=\frac{C_L^2}{\pi AR}.
\end{equation}
The $AR$ of the RTA in RDNP is 12, whereas it is only 5.35 for Avion. Furthermore, washout seems to have an effect on the possible $C_{D_{i}}$ reduction. In the present case washout is zero and, as earlier observed, for a wing with higher washout (case 1 vs. case 8 of RDNP, washout increased from $3^\circ$ to $5^\circ$), the $C_{D_{i}}$ reduction from the respective control wing increased from $9.6\%$ to $35.9\%$, even though the $C_{D}$ was optimised in the latter. This is because with greater washout the circulation distribution on the wing outside the propeller span\textendash station is farther from elliptic/ optimal.
\paragraph{\underline{$C_{D}$ Optimisation}:} Allowing the $C_{f}$ to be optimised along with $C_{D_{i}}$ in figure \ref{fig:F20} achieves a larger $C_{D}$ reduction of $5.5\%$. Lower $C_{f}$ is directly related to the increased twist along the span in\textendash board of $y/s\approx0.6$ in figure \ref{fig:F20}. Then lowering the twist in the out\textendash board region, along with additional downwash by the propeller, allows less lift ($c_l$) from this part to be produced; and hence circulation distribution is closer to elliptic in figure \ref{fig:F20}. This allows $C_{D_{i}}$ to be relatively unaffected. The chord distribution reduces immediately behind the propeller as it is capable of producing enough lift at smaller chord due to larger streamwise velocity. Chord increases out\textendash board of the propeller which is outside the region of higher streamwise velocity generated by the propeller.

\subsection{Approach 2: Floating $C_L$ within a specific range}\label{sec:approach2}
 \subsubsection{Up to 10\% change in $C_L$ allowed:} 
 One of the key features of this aerofoil in the operating Reynolds range is that it has a long low drag region over a range of aerofoil angle of attack where drag changes very little (figure \ref{fig:F12_1}). In the above simulations one of the constraints was to keep the wing lift coefficient $C_L$ constant, thus restricting any movement along this low drag region. Here, $C_L$ is allowed to vary by up to 10\% on either side, i.e. for Avion $(C_{L,\text{design}}=0.7)$ $C_L$ is allowed to float between $0.63$ and $0.77$. As Avion is designed for surveillance, an improvement in endurance factor as a whole would be beneficial, unless it changes handling qualities, payload capability, or any other significant performace parameter.
 	
\begin{figure}[h!]
 	\centering 
 	\subfloat[$W_3$: Case 3($C_{D_{i}}$ optimisation)- $\Delta C_L=-10\%,$ $\Delta C_{D_i}=19.8\%$, $\Delta C_f=-12.7\%$, $\Delta C_D=-1.6\%$, $\Delta (C_L/C_D)=-11.4\%$]{\includegraphics[width=0.45\textwidth]{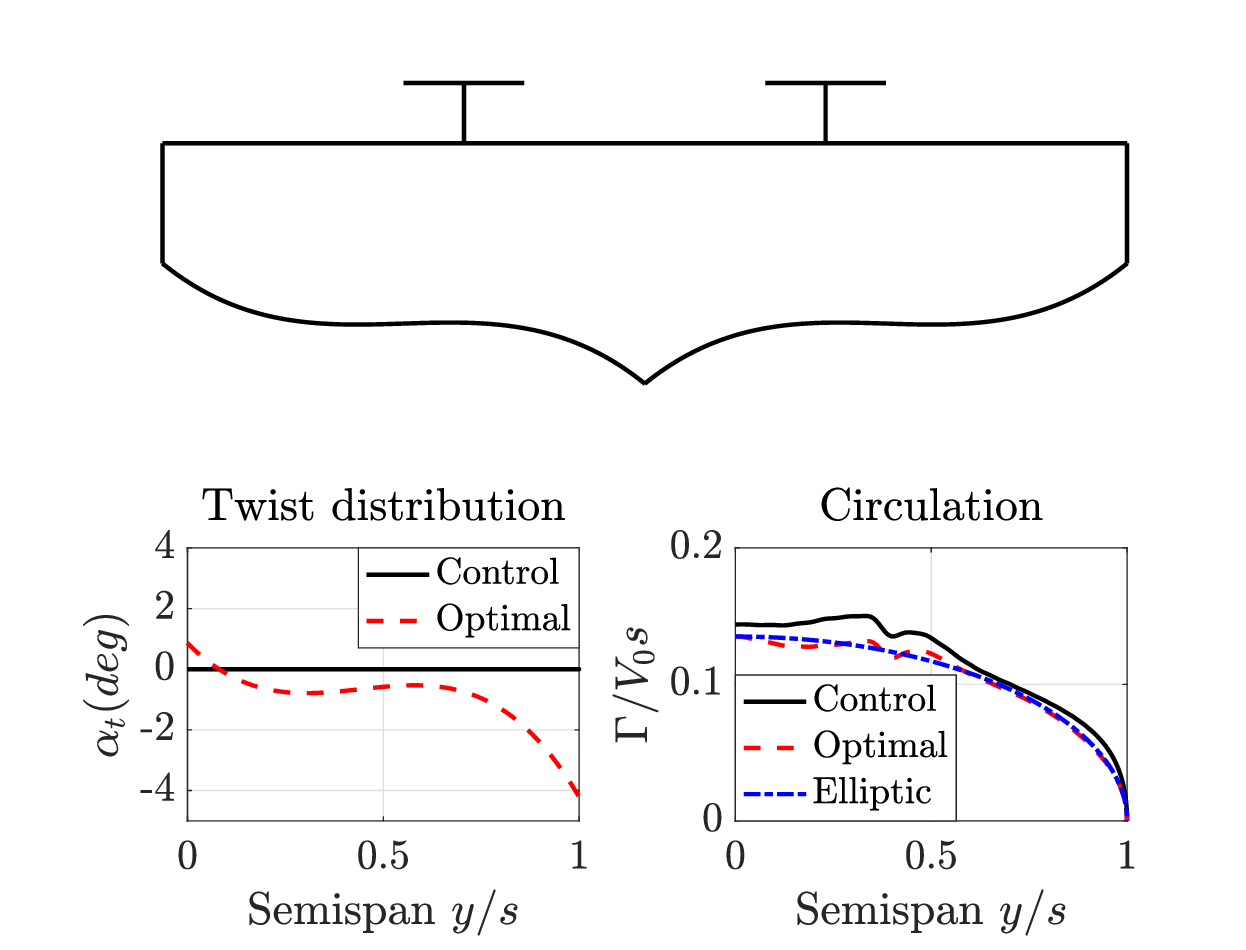} \label{fig:F23}}\\    
 	\subfloat[$W_4$: Case 4($C_{D}$ optimisation)- $\Delta C_L=+10\%$, $\Delta C_{D_i}=-20.7\%$, $\Delta C_f=21.7\%$, $\Delta C_D=7.2\%$, $\Delta (C_L/C_D)=18.6\%$]{\includegraphics[width=0.45\textwidth]{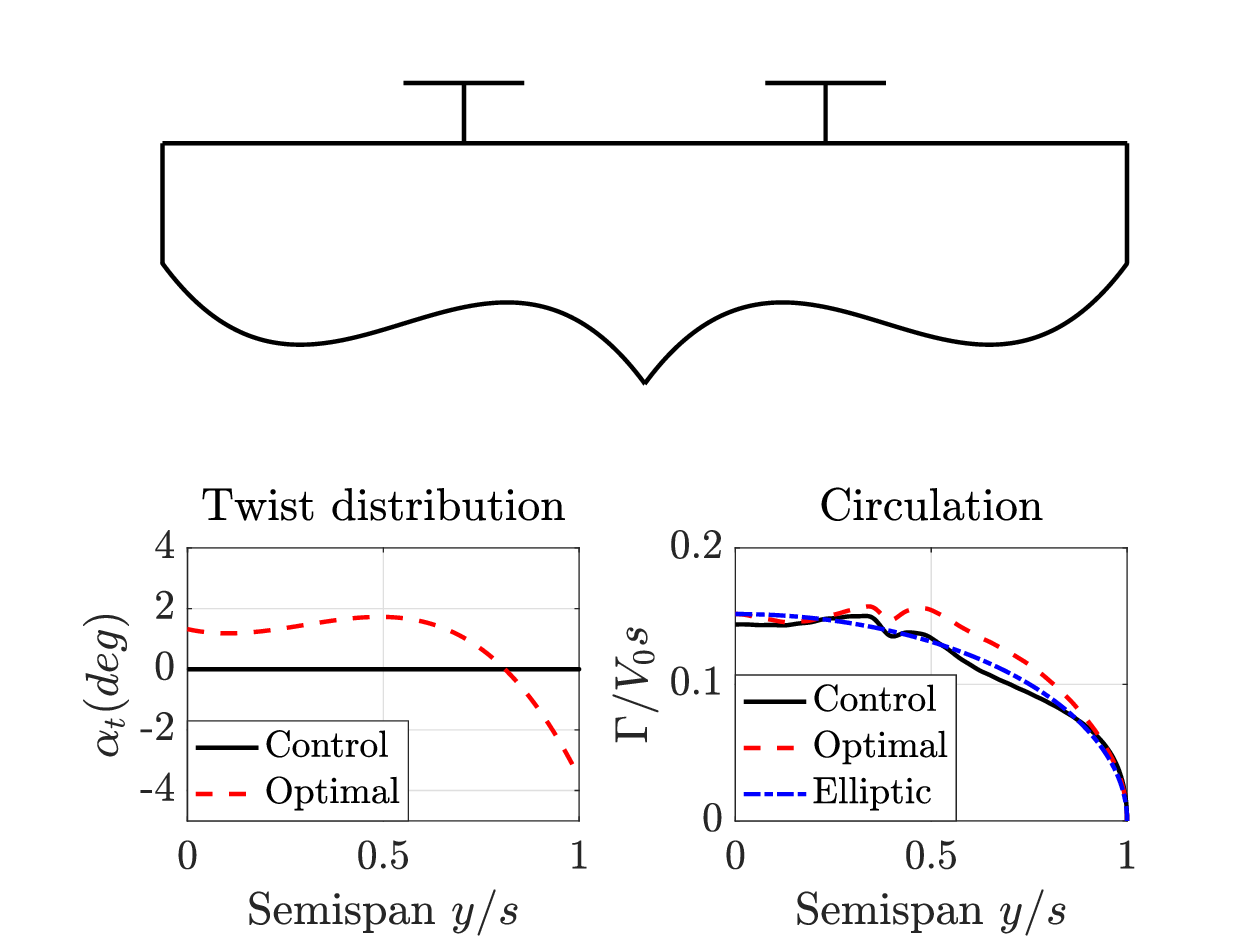} \label{fig:F24}}\\
 	\centering 		
 	\caption {Planforms optimised with $\pm10\%$ variable $C_L$ with initial $C_L=0.7$ @  $\alpha_\text{geo}=-2.25^\circ$\label{2ndOpt}}
 \end{figure}

 To get the initial lift coefficient equal to the design value of 0.7, the angle of attack of the wing is lowered to $-2.25^\circ$ as in the fixed lift case. The change in lift coefficient is defined as \begin{equation}\Delta C_L=C_{L,\text{opt}}-C_{L,\text{NAL}},\end{equation}
 and the change in endurance factor $C_L/C_D$ is given by
 \begin{equation}
 \Delta\Bigg(\frac{C_L}{C_D}\Bigg)=\Bigg(\frac{C_{L}}{C_{D}}\Bigg)_\text{opt}-\Bigg(\frac{C_{L}}{C_{D}}\Bigg)_\text{NAL},
 \end{equation}
 where $C_{L,\text{NAL}}$ and $C_{L,\text{opt}}$ are the lift coefficient of the original and optimised wings respectively. Similar to eq. \ref{eq:drag_change_eq} these definitions allow a favourable change to be represented by a positive number.
 Figure \ref{2ndOpt} shows the optimisation details in terms of shape and circulation profiles and table \ref{table:Table_3} summarises the results of the optimisation. 
	\begin{table*}
\centering
\renewcommand{\arraystretch}{1.5}	
\begin{tabular}{C{3.2cm}||C{2.5cm}|C{2.5cm}|C{2.5cm}|C{2.5cm}|C{2.5cm}} 
 \bfseries{Optimisation Parameter} &\bfseries{ \boldmath$\Delta C_{D_i}$ $(counts/\%)$ (Ini. $C_{D_i}$= \num{0.02384})}&\bfseries{ \boldmath$\Delta C_f$  $(counts/\%)$ (Ini. $C_f$= \num{0.0459})}& \bfseries{ \boldmath$\Delta C_D$  $ (counts/\%)$ (Ini. $C_D$= \num{0.0698})}&\bfseries{ \boldmath$\Delta C_L$  $ (counts/\%)$ (Ini. $C_L$= \num{0.70101})}&\bfseries{ \boldmath$\Delta (C_L/C_D)$ (\%)}\\  \hline \hfill 	
 \bfseries{$C_{D_i}$ Figure \ref{fig:F23}- Case 3}&47.3/ 19.8\% &-58.2/ -12.7\% &-10/ -1.6\% &-701/ -10\% &-11.4\%  \\	
 \bfseries{$C_{D}$ Figure \ref{fig:F24}- Case 4}&-49.3/-20.7\% &99.9/ 21.7\% &50.6/ 7.2\% &701/ 10\% &18.6\%  \\	
 \bfseries{$C_{L}/C_D$ Case 5}&-49.3/-20.7\% &99.9/ 21.7\% &50.6/ 7.2\% &701/ 10\% &18.6\%  \\	
 \end{tabular}
 \caption{Avion optimisations with variable $C_L$; 1 count= \num{0.0001} (Aircraft angle of attack =$-2.25^\circ$)\label{table:Table_3}} 
\end{table*}
 \paragraph{\underline{$C_{D_{i}}$ Optimisation}}: Optimising for $C_{D_{i}}$ leads to a 20\% reduction in $C_{D_i}$ by reducing the twist in the outboard part of the wing (which reduces the lift in this part and hence makes the circulation distribution closer to elliptic\textemdash figure \ref{fig:F23}). However, this reduction in twist to large negative angles (twist distribution in figure \ref{fig:F23}) increases the sectional drag in the outboard part of the wing. This occurs because the  aerofoil sections are further to the left of the low drag range of figure \ref{fig:F12_1} and it leads to an increase in $C_f=D_P/(1/2\rho V_\infty^2S)$ (equation \eqref{Eq:DragEquations}). In this case increase in $C_{f}$ is larger than the improvement in $C_{D_{i}}$ such that a small increase (1.6\%) in $C_{D}$ is observed.
 \paragraph{\underline{$C_{D}$ and Endurance Optimisation}}: Optimising for $C_{D}$ increases $C_{D_{i}}$ but decreases $C_{f}$, resulting in an overall $C_{D}$ benefit of about $7\%$. However, the lift coefficient is increased by the maximum permitted $10\%$, i.e. $C_L=0.77$. This improves the endurance factor by $18.6\%$, which should translate to increased range according to the range equation for battery powered aircraft given by \cite{ref15}
 \begin{equation} \label{Eq:Range_Eq}
 R=E^*\eta_\text{total}\frac{1}{g}\frac{L}{D}\frac{m_\text{battery}}{m_\text{total}},
 \end{equation}
 where $E^*$ (measured in Wh/kg) is the specific energy capacity of the battery, $\eta_\text{total}$ is the total system efficiency i.e. from battery to propulsive power, $\eta_\text{total}=P_\text{propulsive}/P_\text{battery}$ and $g$ is the acceleration due to gravity. Optimisation is obtained by changing the twist distribution in figure \ref{fig:F24}. Twist is increased to an average value of about $+1.5^\circ$ (figure \ref{fig:F24}) along most of the span. However, to bound the increase in $C_{D_{i}}$ the twist is made negative in the out\textendash board part of the wing (beyond $y/s\approx 0.8$).
 
 This also optimises the endurance parameter, $C_L/C_D$, as can be noted by comparing the parameters of case 4 ($W_4$) and 5 ($W_5$) in table \ref{table:Table_3}, where optimising specifically for endurance factor in the latter leads to the same results. Hence, the planform of $W_5$ (same as that of $W_4$) is not shown separately.
 		
 Initially the aerofoil sections along the wing lie near the left end of low drag region in the aerofoil drag polar (figure \ref{fig:F12_1}). If the restriction on $C_L$ is slightly relaxed, the aerofoils can be smoothly twisted along the span in a manner such that higher $C_L$ can be obtained for lower $C_f$. The increase in $C_{D_{i}}$ level that is the penalty of higher $C_L$ (eq. \ref{eq:cdi}) is more than compensated by the $C_{f}$ benefits. This allows an increase in $L/D$ ratio that directly increases the range (eq. \ref{Eq:Range_Eq}). 
 
 Allowing the $C_L$ to vary implies that for the same weight of the aircraft the cruise speed must be changed. Hence, the propeller slip\textendash stream that forms an input to the optimiser should be changed, by coupling the slip\textendash stream measured in an experiment or evaluated from a numerical solver at a different advance ratio $J$. This is because, with a corresponding drop in $J$, thrust is increased (\emph{cf.} figures \ref{fig:F2} and \ref{fig:F5}). However, as only a $3.2\%$ decrease in streamwise\textendash speed is required to obtain the same lift force during cruise (for a $10\%$ increase in lift coefficient), the change in the propeller slip\textendash stream would be small. Therefore, the slip\textendash stream is not coupled to the optimiser here.
 
\subsubsection{Up to 50\% change in $C_L$ allowed:}
\begin{figure}[h!]
	\centering 
	\includegraphics[width=0.45\textwidth]{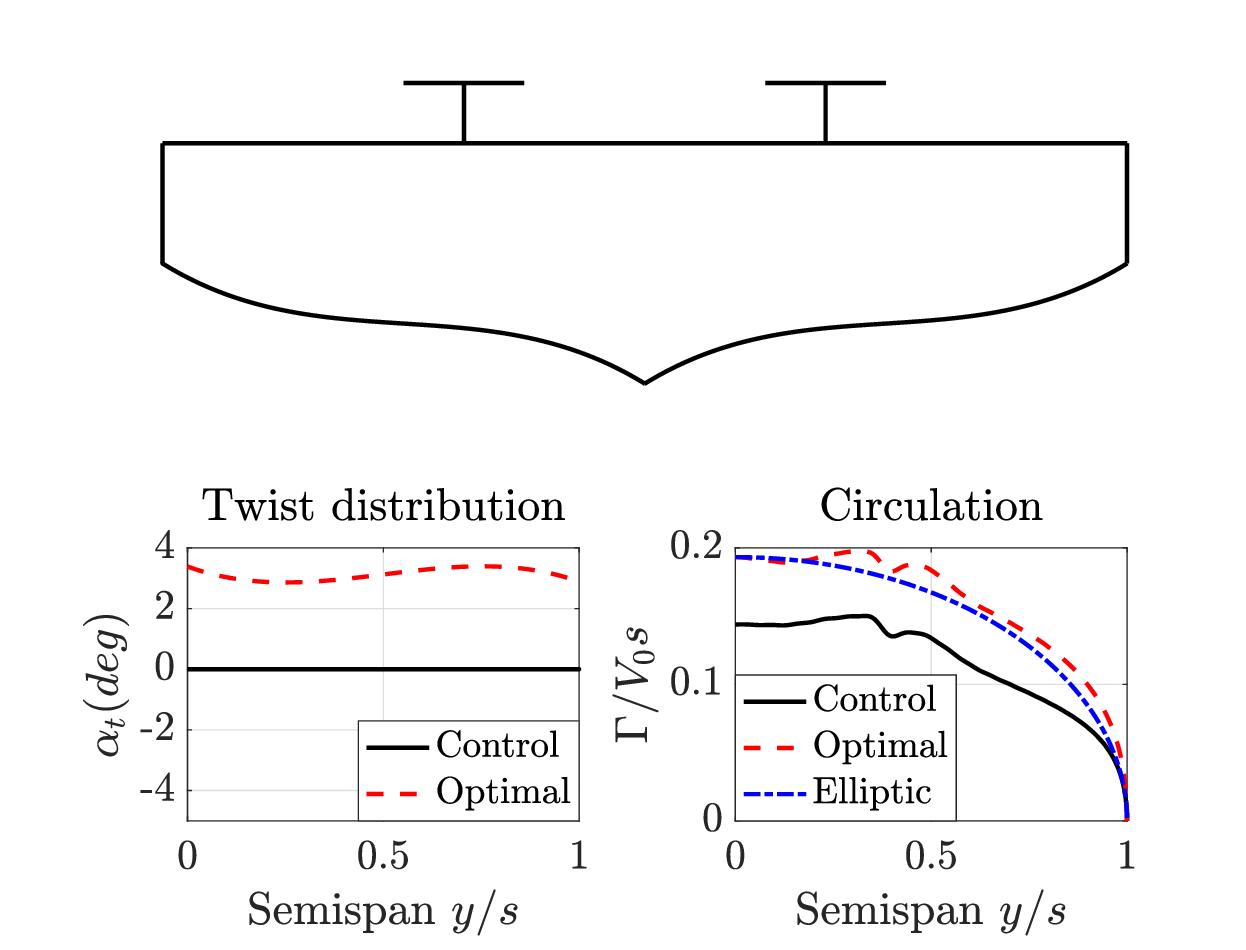} 
	\caption {Planform optimised for endurance with $\pm50\%$ variable $C_L$ with initial $C_L=0.7$ @ $\alpha_\text{geo}=-2.25^\circ$.  Initial drag values: $C_D= \num{0.0698}$, $C_{D_i}=$ \num{0.0238}, $C_f=$ \num{0.0459}.$W_{6}$: Case 6- $\Delta C_L=37.1\%$, $\Delta C_{D_i}=-87.1\%$, $\Delta C_f=47.4\%$, $\Delta C_D=1.5\%$, $\Delta (C_L/C_D)=39.2\%$\label{fig:F29}}
\end{figure}

The permissible change in $C_L$ was further relaxed to $50\%$ and the wing was optimised for endurance only. The results are shown in figure \ref{fig:F29}. An optimised wing planform (wing $W_6$) is obtained such that there is a 39.2\% increase in the endurance factor. But this would require the aircraft to cruise at 37.1\% higher $C_L$, thus forcing the aircraft to be slowed down by about 6.1\% of its original speed, if the same amount of payload has to be carried. 
\vspace{-0.2in}
\subsection{Features of optimised designs}\label{sec:interpret_results}
We recall that the propeller is positioned at 30\% along the wing span from the root (table \ref{table:Table_1} and fig. \ref{fig:F1_4} and it is rotating upward in\textendash board). In all the optimised wings (figures \ref{Fixed_CL_opt} to \ref{fig:F29}) it can be noticed that the general trend for the modified chord distribution is to reduce the chord immediately behind and in\textendash board of the propeller while smoothly increasing it out\textendash board (as in RDNP). Additionally, the trend in the optimised twist distribution in $W_1$ to $W_5$ is to have larger twist ($\alpha_\text{twist}$) out\textendash board and smaller twist in\textendash board of the propeller axis i.e a negative washout. This is consistent with the observation of RDNP where a higher drag benefit was obtained for the original wing with a higher washout. The wing downwash effect is larger near the tip, and also the wing is twisted to negative angles in this region. This implies that the optimised wings may be vulnerable to tip stall on the pressure side. Also the results may be affected by extrapolation of aerofoil characteristics in figure \ref{fig:F12}. It may be worthwhile to check this with a higher fidelity method such as a RANS solver. Unlike RDNP, the circulation distribution of the optimised profiles with most drag benefit is not elliptic as the $C_f$ is not only initially larger but it can be optimised more efficiently due to the laminar drag bucket region of figure \ref{fig:F12_1}.
\begin{figure}[h!]
	\centering 
	\subfloat[Lift Curve]{\includegraphics[width=0.49\textwidth]{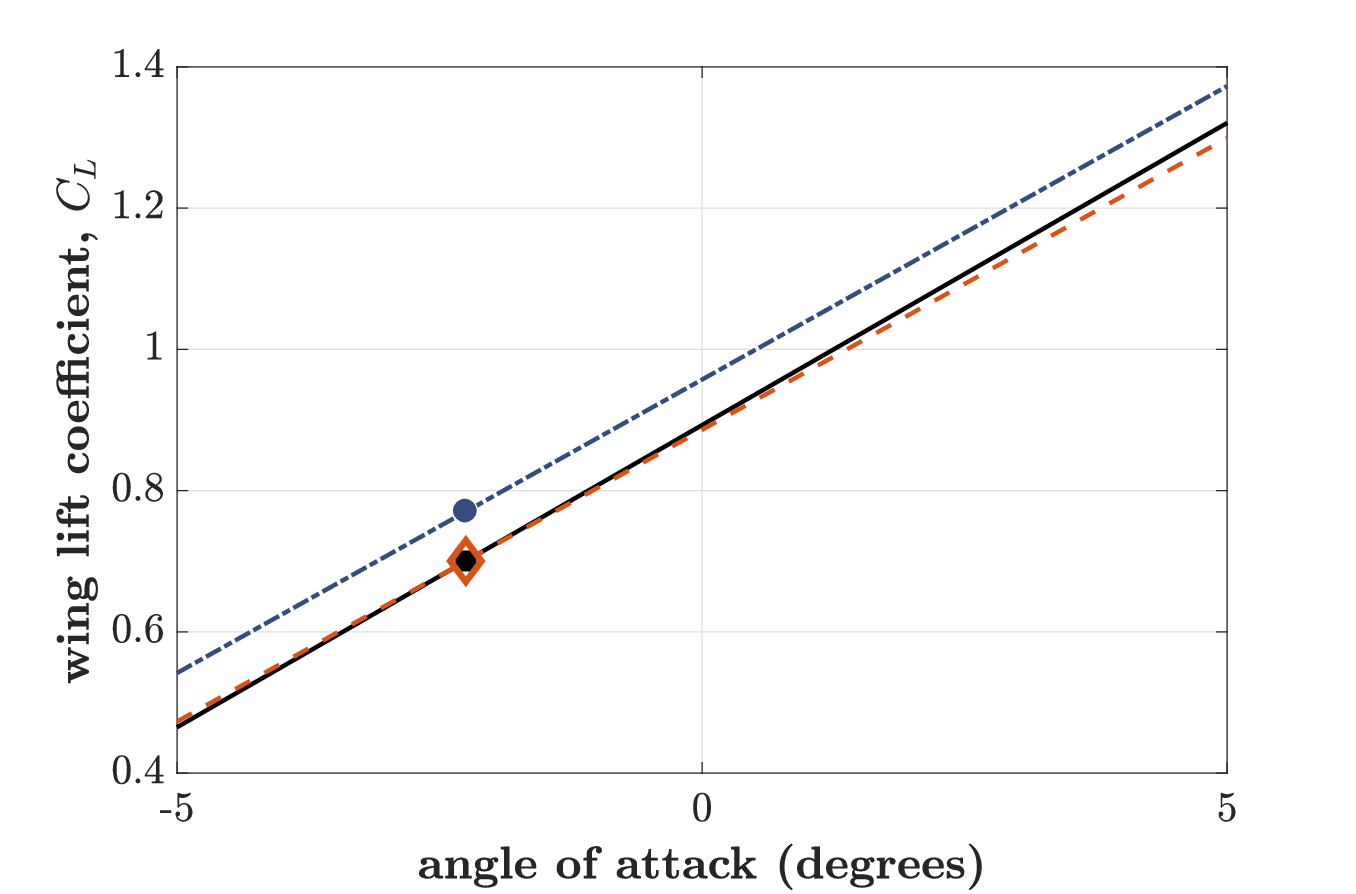}
		\label{fig:compare_cl}}\\
	\centering    		
	\subfloat[Drag Polar]{\includegraphics[width=0.49\textwidth]{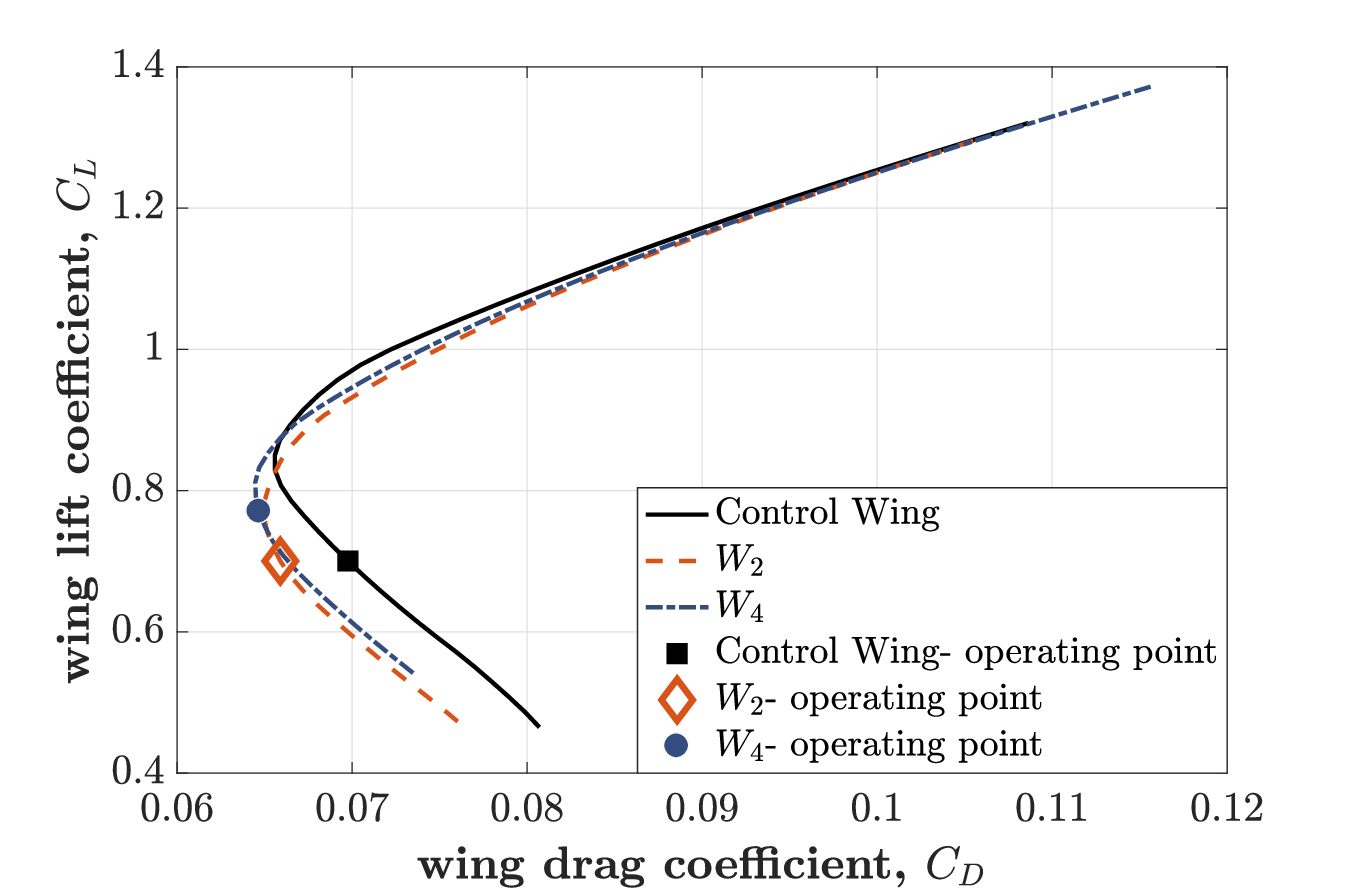} \label{fig:compare_cd}}
	\caption {Comparison of the wing polars and the operating points of the control wing, $W_2$ and $W_4$. The legends for the curves is same for both the plots.\label{fig:compare_polars}}
\end{figure}

Comparing the wing polars of the control wing $W_0$ and the optimised wings $W_2$ and $W_4$, figure \ref{fig:compare_polars} reveals that even when the endurance factor is not explicitly specified as the cost function ($W_4$), variable $C_L$ optimisation attempts to find the optimal $C_L/C_D$ for this Avion design evaluated without considering the propeller slip\textendash stream \cite{NAL_Personal}. For the original operating $C_L=0.7$, optimising for $C_{D}$ ($W_2$) lowers the wing lift slope (figure \ref{fig:compare_cl}) and also reduces the drag (figure \ref{fig:compare_cd}) by shifting the drag polar left in the region of operating $C_L$. However, allowing the operating $C_L$ to change permits the full potential of this optimisation to be realised. Relative to the original wing the drag polar is shifted to the left for a slightly larger range of wing angle of attack between $-5^\circ$ and $5^\circ$. The operating point on the drag polar of $W_4$  in figure \ref{fig:compare_cd} is also near the optimal location. The small difference from the exact optimal location is because the $C_L$ variation is restricted to 10\%. The lift curve is shifted up in figure \ref{fig:compare_cl} because the wing twist is on average 1.5$^\circ$ along most of the wing span in figure \ref{fig:F24}.

\subsection{Changing the design point or geometry of control wing}
The general trends described above along with the observations from the wing polars in figure \ref{fig:compare_polars} can provide useful suggestions for further changes to the preliminary control wing specified in table \ref{table:Table_1}. Such ideas are explored below.
\subsubsection{Altering operating conditions}
\begin{figure}[h!]
	\centering 
		\includegraphics[width=0.45\textwidth]{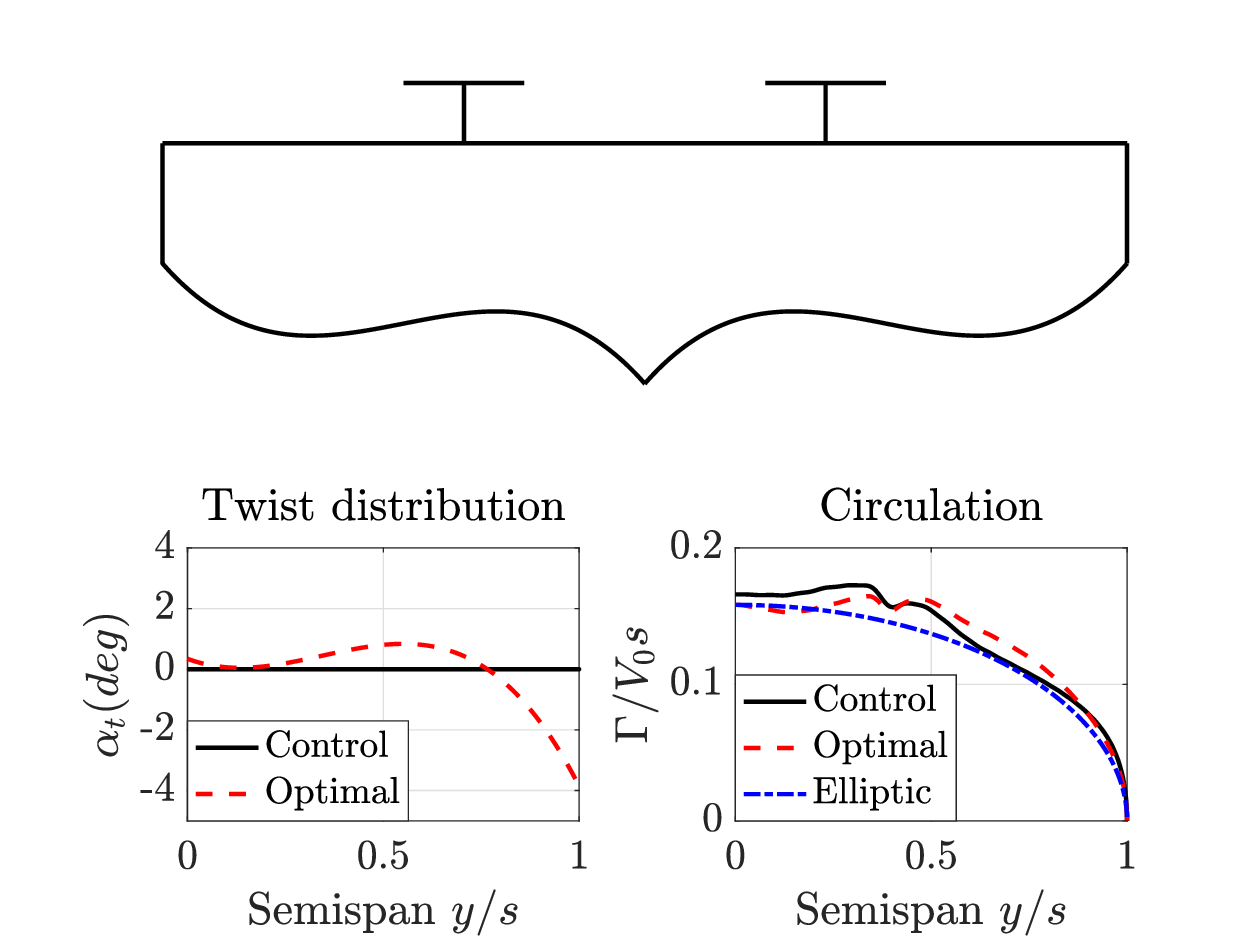} 
	\caption {$W_7$: Case 7- Wing planform with original geometry in figure \ref{fig:Orig_wing} as control wing optimised for $C_{D}$ at constant lift coefficient, $C_L=0.81$ @ $\alpha_\text{geo}=-1^\circ$. Initial drag coefficient values: $C_D= \num{0.0659}$, $C_{D_i}=$ \num{0.0316}, $C_f=$ \num{0.0343}. $\Delta C_{D_i}= 0.25\%,\ \Delta C_f=4.5\%,\ \Delta C_D=2.5\%$\label{New_Wing}}
\end{figure}

A possible next step in the overall design iteration would be to rethink the operating $C_L$ and geometric angle of attack of the wing. As a first attempt the operating $C_L$ is changed to be nearer to the optimal of the control wing polar of figure \ref{fig:compare_cd}. The wing geometric angle of attack is changed from $-2.25^\circ$ to $-1^\circ$ such that the operating $C_L$ and $C_D$ are now $0.81$ and \num{0.0659} respectively i.e. 5.6\% less $C_{D}$ with 16\% higher $C_{L}$ compared to the original control wing specified in table \ref{table:Table_1}. The results for the optimisation carried out for $C_{D}$ with a fixed wing $C_L=0.81$ in this new configuration are shown in figure \ref{New_Wing}. 

In this case, the changes in the wing twist and chord distribution follow the general trend described in section \ref{sec:interpret_results}. An improvement in $C_f$ leads to a $C_D$ improvement of 2.5\%. The chord and twist distributions are similar to those obtained for $W_4$ in figure \ref{fig:F24}, where the original control wing was optimised for $C_{D}$ with 10\% variation in $C_L$ allowed. Considering the difference in $\alpha_\text{geo}$ between the two wings i.e. $-2.25^\circ$ for $W_4$ and $-1^\circ$ for $W_7$, these are two different locations near the optimal $C_L/C_D$ on very similar wing polars (cf. figure \ref{fig:compare_cd} for $W_4$). The $C_L$ and $C_D$ for $W_4$ are $0.77$ and \num{0.0647} respectively, whereas for $W_7$ the corresponding values are $0.81$ and \num{0.0643}. 

\subsubsection{Changing control wing washout}
\begin{figure}[h!]
	\centering 
	\includegraphics[width=0.45\textwidth]{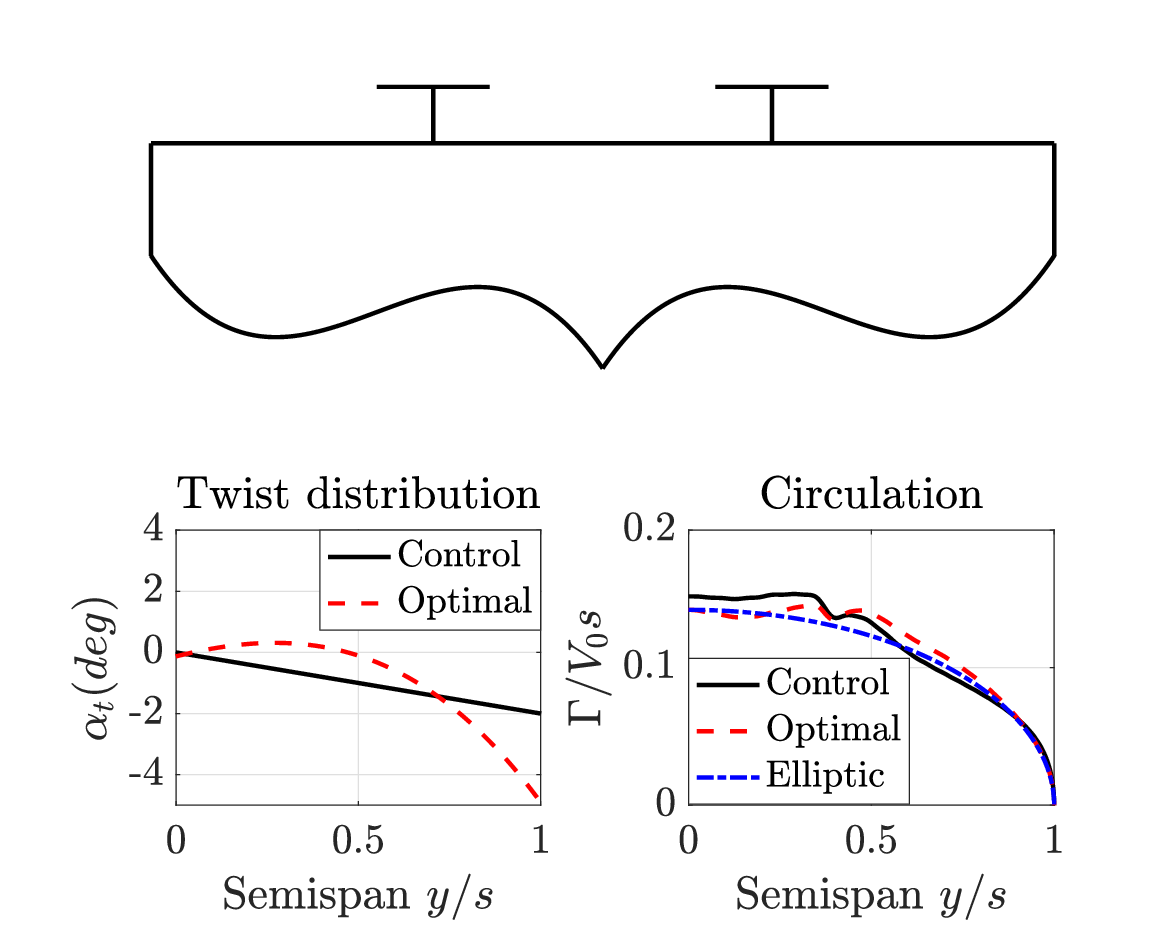} 
	\caption {$W_8$: Case 8- Wing planform with washout changed from $0^\circ$ in the original wing in figure \ref{fig:Orig_wing} to $-2^\circ$. $C_{D}$ optimisation at constant lift coefficient, $C_L=0.7$ at $\alpha_\text{geo}=-1.4^\circ$.Initial drag values: $C_D= \num{0.0697}$, $C_{D_i}=$ \num{0.0239}, $C_f=$ \num{0.0458}. $\Delta C_{D_i}= 1.2\%,\ \Delta C_f=7.6\%,\ \Delta C_D=5.4\%$\label{New_Wing2}}
\end{figure}

Following the general trend of a negative washout in the optimised profiles of $W_1$ to $W_5$ another possible step could be to introduce a negative washout in the the original wing with linear spanwise chord distribution. Hence, the original wing in figure \ref{fig:Orig_wing} with specifications mentioned in table \ref{table:Table_1} is altered to have a linear variation in twist from $0^\circ$ at the wing root to $-2^\circ$ at the wing tip. According to LLT calculations the specified $C_L=0.7$ for this wing is obtained at a more horizontal wing geometric angle of attack $\alpha_\text{geo}=-1.4^\circ$ compared with the original wing ($\alpha_\text{geo}=-2.25^\circ$). At these operating conditions ($C_L=0.7$) the $C_D$ of this new control wing is almost the same as that for the original control wing specified in table \ref{table:Table_1}, i.e. now $C_D=$ \num{0.0697} compared to \num{0.0698} for the latter. The results for optimising this wing for $C_D$ whilst keeping fixed $C_L=0.7$ are shown in figure \ref{New_Wing2}. This optimised wing, referred to as $W_8$, is similar to $W_2$ obtained by optimising the original wing for $C_D$ (figure \ref{fig:F24}). $W_8$ has almost the same drag coefficient ($C_D=0.2\%$ lower than $W_2$) and exactly the same $C_L$ ($=0.7$) as $W_2$. However $W_8$ achieves this $C_L$ at $\alpha_\text{geo}=-1.4^\circ$ whereas $W_2$ requires $\alpha_\text{geo}=-2.25^\circ$.
\subsubsection{Aerofoil selection}
Perhaps there is an indication towards having different aerofoil sections at different spanwise locations along the span of the Avion wing. As $\alpha_\text{geo}$ was negative for all the wings analysed here, an aerofoil with an optimal $c_l/c_d$ at lower angle of attack than E423 would be more suitable for Avion if the design $C_L$ is to be kept at the initially specified $0.7$. Furthermore, compared to their respective control wing the twist is increased in the in\textendash board sections and reduced near the tip in all of the optimised geometries with improved parameters, excepting $W_6$. Here variation in aerofoil section near the tip may be further tuned to have an aerofoil with an even lower optimal angle of attack with lower corresponding lift. For example an aerofoil with a lesser camber than E423 in figure \ref{fig:e423shape} can be used. If these optimally chosen aerofoils also possess an appreciable low drag region, an initial wing geometry operating in this range can be further optimised for $C_{D}$ with variable $C_L$ as demonstrated in section \ref{sec:approach2}.

\vspace{0.2in}

It would be highly beneficial to test these suggestions based on LLT using interpolation of aerofoil characteristics in figure \ref{fig:F12} with a higher fidelity tool such as RANS, and also if possible with wind tunnel tests.

\section{Conclusions}\label{sec:Conclusions}
In this paper, the methodology developed by RDNP is extended to optimise wings of a tractor\textendash propeller MAV, Avion. The capability of an open source toolbox (OpenFOAM) to calculate the incompressible propeller slip\textendash stream in case of Avion is demonstrated. This would replace the compressible Euler code (Prop-EULER) of RDNP in case the experimental data are not available and propeller geometry in known. 

In the actual wing optimisation process, the cost function used here is different from that used in RNDP; so are the Reynolds number, aerofoil sections, and operating conditions. Also, the $C_{D_{i}}$ for the smaller aircraft is much higher in absolute terms. Hence, an equal or even larger improvement in the absolute value of the drag coefficient ($C_D$ counts) is much lower than the RDNP in percentage terms. For a constant wing $C_L$, even when the $C_{D_{i}}$ is optimised, most of the benefit arises from the $C_f$ improvement due to the relatively large low drag range of the E423 aerofoil (figure \ref{fig:F12_1}). Higher benefits are obtained by optimising $C_{D}$.

Due to this property of the E423 aerofoil used in Avion, a $C_f$ improvement is achievable that can be fully realised if the operating lift coefficient is allowed to vary even by as little as 10\%. There is a potential for reducing $C_D$ whilst increasing $C_L$ under the current operating conditions, leading to an improvement in the $L/D$ ratio. For an MAV being designed for surveillance this performance parameter is of utmost importance, as it would amount to a significant improvement in range for a small change in lift force or cruise velocity. In this particular case the reduction in $C_f$ more than compensates for the increase in $C_{D_{i}}$ that accompanies the increased lift, resulting in an improvement in endurance factor by $18.6\%$ in one of the optimised solutions (wing $W_4$).

The optimisation philosophy used here has the capability to find the optimal operating conditions if certain aspects can be ignored in the initial design iteration. In the initial Avion design, the propeller slip\textendash stream effects are often ignored \cite{NAL_Personal}. Allowing for a variable $C_L$ in the optimisation process allows better operating conditions to be obtained once the propeller effect is added. From the wing polar obtained from lifting line theory for the control wing with the propeller effect added, the optimal $L/D$ ratio for Avion is obtained for $C_L=0.85$ at $\alpha=-1.3^\circ$. This suggests reconsidering the value of the design $C_L$, initially proposed to be 0.7. The observation of twist distributions of the optimised profiles obtained from the original wing suggests a negative washout could be beneficial for this wing. Such new directions can be constructive in general during the preliminary design stage.

\section{Acknowledgement}

We would like to thank Sh. Shyam Chetty, Director, NAL, Bengaluru and Dr. G. N. Dayananda, Chief Scientist (CSMST)
 for supporting this project, Dr. G. Ramesh, former head of MAV Unit and his colleagues for providing data on Avion design and Mr. Shashank Anand for performing wind tunnel tests to quantify the propeller slip\textendash stream. AS is grateful for the opportunity to discuss various issues cropping up in this project with Milind Dhake and B.R. Rakshith.

\bibliographystyle{vancouver}
\bibliography{Paper}

\begin{thebibliography}{1}

\bibitem{Rakshith2015}
Rakshith BR, Deshpande SM, Narasimha R, Praveen C.
\newblock {Optimal Low-Drag Wing Planforms for Tractor-Configuration
  Propeller-Driven Aircraft}.
\newblock Journal of Aircraft. 2015 April;\ 52(6):1791-801.
\newblock Available from: \url{http://dx.doi.org/10.2514/1.C032997}.

\bibitem{Drela1989}
Drela M.
\newblock XFOIL: An analysis and design system for low Reynolds number
  airfoils.
\newblock In: Low Reynolds number aerodynamics. Springer; 1989. p. 1-12.

\bibitem{ref15}
Hepperle M.
\newblock Electric Flight- Potential and Limitations. NATO-OTAN MP-AVT-209-09;
  Oct 2012.
\newblock Available from:
  \url{http://www.mh-aerotools.de/company/paper{\_}14/MP-AVT-209-09.pdf}.

\bibitem{Wahono2013}
Wahono S.
\newblock {Development of Virtual Blade Model for Modelling Helicopter Rotor
  Downwash in OpenFOAM}.
\newblock DSTO-TR-2931, Aerospace Division, Australian Defence Science and
  Technology Organisation; 2013.
\newblock Available from:
  \url{http://dspace.dsto.defence.gov.au/dspace/bitstream/dsto/10520/1/DSTO-TR-2931
  PR.pdf}.

\bibitem{Deters2014}
Deters RW, Ananda GK, Selig MS.
\newblock Reynolds number effects on the performance of small-scale propellers.
\newblock In: Proceedings of the 32nd AIAA Applied Aerodynamics Conference,
  Atlanta, GA, USA; 2014. p. 16-20.

\bibitem{Aerofoil_summary}
Selig M, Lyon C, Giguere P, Ninham C, Guglielmo JJ.
\newblock {Summary of Low-Speed Airfoil Data, vol. 2}.
\newblock SoarTech Publications, Virginia Beach, Virginia; 1996.
\newblock Available from:
  \url{http://m-selig.ae.illinois.edu/uiuc_lsat/Low-Speed-Airfoil-Data-V2.pdf}.

\bibitem{Hartman1938}
Hartman EP, Biermann D. {The aerodynamic characteristics of full-scale
  propellers having 2, 3 and 4 blades of Clark Y and R.A.F 6 airfoil sections}.
  NACA Report 640; 1938.
\newblock Available from:
  \url{http://naca.central.cranfield.ac.uk/reports/1938/naca-report-640.pdf}.

\bibitem{Propeller_Database}
Krishnan A, Kumar G. {UIUC Propeller Database- Volume 1}; 2015.
\newblock Available from:
  \url{http://m-selig.ae.illinois.edu/props/volume-1/propDB-volume-1.html}.

\bibitem{NAL_Personal}
Shashant A, Ramesh G. APC 11 X 7 thin electric propeller PIV experiment and
  Avion CAD, operating conditions and CFD results, CSIR-NAL, Personal
  Communication; 2015.

\end{thebibliography}
\end{document}